\documentstyle[12pt]{article}
\newcommand{\be}{\begin{equation}}
\newcommand{\ee}{\end{equation}}

\begin{document}
\begin{titlepage}
\sloppy
\thispagestyle{empty}

\mbox{}
\hspace{10cm} {\Large Preprint ITEP  No.62--95}

\vspace*{5cm}
\begin{center}
{\LARGE \bf Nucleon Spin Structure. Sum Rules. } \\

\vspace{2mm}
%{\LARGE\bf title continued e.g.}\\

\vspace{2mm}
\large
B.L.Ioffe %Your Name
\\
\vspace{2em}
{\it  Institute of Theoretical and\\
Experimental Physics
%Your Institution,
 \\
B.Cheremushkinskaya 25, 117259 Moscow, Russia}

e-mail: ioffe@vxdesy.desy.de, ioffe@vxitep.itep.ru
%\today

\vspace{5mm}
Lectures at the International School of Nucleon\\
Structure, Erice, 1995
\end{center}
%\vspace*{\fill}
%\end{document}
\vspace{3cm}
%\begin{abstract}
%\noindent
\underline{\bf Contents}

\vspace{3mm}
1. The sum rules or $\Gamma_{p,n}$.Theoretical status.

2. Calculations of matrix elements over the polarized nucleon by the QCD
sum rule approach.

3. Twist-4 corrections to $\Gamma_{p,n}$ from QCD sum rules.

4. Gerasimov, Drell-Hearn (GDH) sum rules. The estimate of higher twist
corrections to

$\Gamma_{p,n}$
using interpolation between $\Gamma_{p,n}$ and GDH sum rules.

5. Comparison with experiment.

6. The calculations of polarized structure functions by the QCD sum rule
method.

7. Calculation of the chirality violating structure function $h_1(x)$ by QCD
sum rules.

8. Conclusions.
\end{titlepage}
%\end{document}
%\end{document}

\newpage
\section{The sum rules for $\Gamma_{p,n}$. Theoretical Status.}

%\vspace{1.5mm}
I start with the consideration of the sum rules for the first
moments of the spin structure functions $g_{1p,n}(x,Q^2)$
%\end{document}
\be
\Gamma_{p,n} (Q^2) = \int \limits_0 ^1 dx g_{1 p,n} (x, Q^2)
\ee
I will discuss the uncertainties in the theoretical predictions for
$\Gamma_{p,n}$ and compare the theoretical expectations with experimental
data. The aim of this consideration is to obtain from the experiment the
restrictions on the uncertainties in the theoretical description of the
problem.

Consider first the Bjorken sum rule [1], which now is a corner stone not
only of the problem of nucleon spin structure functions,but of the whole
theory of deep inelastic lepton-hadron scattering in QCD. (It is interesting
to note that in` his original paper $^{[1]}$ in 1966 Bjorken had classified
this sum rule as a "worthless equation").

The Bjorken sume rule reads:
%2
\be
\Gamma_p
(Q^2) - \Gamma_n(Q^2) = \frac{1}{6} g_A [1 - \frac{\alpha_s(Q^2)} {\pi} -
3.6 \left (\frac{\alpha_s(Q^2)}{\pi}\right )^2 - 20 \left
(\frac{\alpha_s(Q^2)}{\pi}\right )^3] + \frac{b_{p-n}}{Q^2},
\ee
where $g_A$
is the axial $\beta$-decay coupling constant and the last term in (2)
represents the twist-4 correction. The perturbation QCD corrections are
known up to the third order $^{[2,3]}$ (there is also an estimate of the
fourth order term ${[^4]}$, which is not included in (2), since the
uncertainties in the included terms are larger than its contribution). The
coefficients in (2) correspond to the number of flavours $N_f = 3$. (In the
domain of existing experiments only three flavours of quarks are effective).
In what follows I will so often transfer the data to the standard reference
point $Q^2_0 = 10.5 GeV^2$ -- the mean value of $Q^2$, at which EMC and
SMC experiments were done.

Let us discuss the perturbative corrections in (2). Today there is a
serious discrepancy in the values of $\alpha_s$ found from different
experiments. The average value of $\alpha_s$, obtained at LEP is ${[^5]}$
%3
\be
\alpha_s (m^2_Z) = 0.124 \pm 0.007
\ee
In two-loop approximation this value corresponds to the QCD parameter
$\Lambda_3$ for three flavours
%4
\be
\Lambda_3 = 430 \pm 100 MeV
\ee
On the other hand,Tte data on the $\Upsilon \rightarrow$ hadrons decay give
$^{[5]}$ (the first $\alpha_s$ correction $^{[6]}$ is accounted)
%5
\be
\alpha_s (m^2_b) = 0.178 \pm 0.010
\ee
from which it follows
%6
\be
\Lambda_3 = 170 \pm 30 MeV
\ee
The small error in (6) is caused by the fact that the partial width
$\Gamma(\Upsilon \rightarrow 3g)$, from which (5) was determined, is
proportional to $\alpha^3_s$ and the $\alpha_s$ correction to it is small. A
strong contradiction of (4) and (6) is evident.

The overall fit $^{[7]}$ of the data of deep inelastic lepton-nucleon
scattering gives in the NLO approximation $\Lambda_3 = 250 MeV$ (the error
is not given). New data in the domain $m^2_Z$ indicate lower
$\alpha_s(m^2_Z)$  (SLD $^{[8]}: 0.118 \pm 0.013, 0.112 \pm 0.004$; OPAL
$^{[9]}$: $0.113 \pm 0.012$), but the data of AMY $^{[10]}$ on $e^+e^-$
annihilation at $\sqrt{s} = 57.3 GeV$ results in $\alpha_s(m^2_Z) = 0.120
\pm 0.005$. Finally, from $\tau$- decay it was obtained $^{[11]}$:
%7
\be
\alpha_s(m^2_{\tau}) = 0.33 \pm 0.03
\ee
%\end{document}
corresponding to
$$
\Lambda_3 = 380 \pm 60 MeV
$$
But the determination of $\alpha_s$ from $\tau$-decay can be criticized
$^{[12]}$ on the grounds that at such a low scale exponential terms in $q^2$
may persist in the domain of positive $q^2 = m^2_{\tau}$ besides the standard
power-like terms in $q^2$ accounted in the calculation.

In such a confusing situation I will consider two options -- of small and
large perturbative corrections. In the first case I will take $\Lambda_3 =
200 MeV$. Then
%8
\be
\alpha_s(Q^2_0) = 0.180 \pm 0.010 ~~~~~~~~~~~~ \Gamma_p(Q^2_0) -
\Gamma_n(Q^2_0) = 0.194
\ee

In the second one $\Lambda_3 = 400 MeV$ and
%9
\be
\alpha_s(Q^2_0) = 0.242 \pm 0.025 ~~~~~~~~~~~~~~~\Gamma_p(Q^2_0) -
- \Gamma_n(Q^2_0) = 0.187
\ee

The twist-4 contribution was disregarded in (8) and (9). There is also a
discrepancy in its value. The value of $b_{p-n}$ was determined in the QCD
sum rule approach by Balitsky, Braun and Koleshichenko (BBK) $^{[13]}$:
%10
\be
b_{p-n} = -0.015 GeV^2
\ee
On the other hand, the model $^{[14-16]}$ based on connection $^{[17]}$ of
the Bjorken sum rule at large $Q^2$ with the Gerasimov, Drell, Hearn sum
rule at $Q^2 = 0$ $^{[18]}$ gives
%11
\be
b_{p-n} = -0.15 GeV^2
\ee
In what
follows I will consider (10) and (11) as two options which correspond to
small (S) and large (L) twist-4 corrections. I will discuss both approaches
of determination of higher twist corrections in more details below. From my
point of view, no one of these approaches is completely reliable and I will
use them in comparison with experiments only as reference points.

I turn now to the sum rules for $\Gamma_p$ and $\Gamma_n$:
%12
\be
\Gamma_{p,n}(Q^2) = \frac{1}{12} \left \{\left [1 - \frac{\alpha_s}{\pi} -
3.6 (\frac{\alpha_s}{\pi})^2 - 20 (\frac{\alpha_s}{\pi})^3\right ] (\pm g_a
+ \frac{1}{3} a_8) + \right.
\ee
$$\left.\frac{4}{3} \left [1 -
\frac{\alpha_s}{3\pi} - 0.55 (\frac{\alpha_s}{\pi})^2\right ] \Sigma \right
\} - \frac{N_f}{18 \pi} \alpha_s (Q^2) \Delta g (Q^2) + \frac {b_{p,n}}{Q^2}
$$

(The $\alpha_s^2$ correction to the singlet part was calculated in
$^{[19]}$).

According to the current algebra $g_A, a_8$ and $\Sigma$ are
determined by the proton (or neutron) matrix elements of flavour octet and
singlet axial currents
%13
\be
-2m s_{\mu} a_8 = <p, s \mid j^{(8)}_{5 \mu}
\mid p,s > ~~~~~ -2ms_{\mu} \Sigma = <p, s \mid j^{(0)}_{5 \mu} \mid p,s>,
\ee
$$ -2ms_{\mu} g_A = <p, s \mid j^{(3)}_{5 \mu} \mid p,s >
{}~~~~~~~~~~~~~~~~~~~~~(13')$$
where
$$j^{(8)}_{5 \mu} = \bar{u} \gamma_{\mu} \gamma_5 u + \bar{d}
\gamma_{\mu} \gamma_5 d - 2 \bar{s} \gamma_{\mu} \gamma_5 s,~~~~~ j^{(3)}_{5
\mu} = \bar{u} \gamma_{\mu} \gamma_5 u - \bar{d} \gamma_{\mu} \gamma_5 d$$

$$
j^{(0)}_{\mu 5} = \bar{u} \gamma_{\mu} \gamma_5 u + \bar{d}
\gamma_{\mu} \gamma_5 d + \bar{s} \gamma_{\mu} \gamma_5 s
$$
In the parton model $g_A,a_8$ and $\Sigma$ are equal to
%14
\be
g_A = \Delta u - \Delta d ~~~~~~~~ a_8 = \Delta u + \Delta d - 2\Delta S
{}~~~~~~~\Sigma = \Delta u + \Delta d + \Delta s,
\ee
where
%15
\be
\Delta q = \int \limits _{0} ^{1} \left [q _+(x) - q_-(x)\right ]~~~~~~~~~~
q = u,d,s
\ee
and $q_{\pm}$ are the quark distributions with the spin parallel
(antiparallel) to the proton spin, which is supposed to be longitudinal
(along the beam). $\Delta g$ in (12) has the similar meaning, but for
gluons, as  $\Delta q$ for quarks
%16
\be
\Delta g(Q^2) = \int \limits _{0} ^{1} dx \left [ g_+ (x,Q^2) - g_- (x,
Q^2)\right ]
\ee
Unlike $g_A, a_8$ and $\Sigma$ which in the approximation used above are $Q^2$
independent and have zero anomalous dimensions, $\Delta g$ anomalous
dimension is equal to --1. This means that
%17
%\begin{eqnarray}
\be
\Delta
g(Q^2)_{Q^2 \rightarrow \infty} \simeq cln Q^2
\ee
The conservation of the projection of the angular
momentum can be written as
%18
\be
\frac{1}{2} \Sigma + \Delta g (Q^2) + L_z
(Q^2) = \frac{1}{2}
\ee
where $L_z$ has the meaning of the orbital momentum
of quarks and gluons.  As follows from (17),(18) at high $Q^2 ~~ L_z(Q^2)$
must compensate $\Delta g(Q^2)$, $L_z(Q^2) \approx -c ln Q^2$. This means
that the quark model, where all quarks are in $S$-state, failed with $Q^2$
increasing.

The gluonic contribution to $\Gamma_{p,n}$ term, proportional to $\Delta~g$
in (12), was calculated in $^{[2,20-23]}$.
There was a wide discussion in the past years if gluonic contribution
$\Delta_g \Gamma_{p,n}$ to $\Gamma_{p,n}$ is uniquely defined theoretically
or is not $^{[23-36]}$. The problem is that gluonic contribution to the
structure functions, described by imaginary part of the forward
$\gamma_{virt}$--gluon sccattering amplitude (Fig.1) is infrared
dependent. Since in the infrared domain the gluonic and sea quark
distributions

\vspace{2cm}

\hspace{10cm}  Fig.1

\footnotesize
\hspace{10cm}  The photon-gluon scattering

\hspace{10cm}  diagrams, the wavy, dashed

\hspace{10cm}  and solid lines correspond to

\hspace{10cm}  virtual photons, gluons and quarks.

\vspace{2cm}
\normalsize
are mixed and their separation depends on the infrared regularization
scheme, a suspicion arises that $\Delta_g \Gamma_{p,n}$ can have any value.
This suspicion is supported by the fact that in the lowest order in
$\alpha_s$ the terms proportional to $lnQ^2$ are absent in $\Delta_g
\Gamma_{p,n}$ and this contribution looks like next to leading terms in
nonpolarized structure functions where such an uncertainty is well known.

In the framework of the operator product expansion (OPE) there is only one
operator -- the singlet axial current $j^{(0)}_{\mu 5}$ -- corresponding to
the first moment of $g_{1p} + g_{1n}$, i.e., $\Gamma_{p+n}$. This fact can
be also used as an argument in the favour that (in the $\alpha_s$ and higher
orders) the separation of the terms proportional to $\Sigma$ and $\Delta~g$
in (12) is arbitrary.

In order to discuss the problem consider the gluonic contribution
$g_{1p}(x,Q^2)_{gl}$ to the proton structure function $g_{1p}(x,Q^2)$,
described by the evolution equation
%19
\be
g_{1p}(x,Q^2)_{gl} = N_f \frac{< e^2 >}{2} \int \limits _{x} ^{1} \frac
{dy}{y} A \left (\frac{x}{y} \right ) [g_+(y,Q^2) - g_-(y,Q^2)],
\ee
where the asymmetry $A(x_1)$ is determined by the diagrams of Fig.1. The
calculation of the asymmetry $A(x_1), x_1 = -q^2/2pq$ ~results in appearance
of integrals
%nn
$$
\int~ \frac{d^2 k _{\bot}}{(k^2_{\bot} - x_1(1 - x_1)p^2 + m^2_q)^n}, ~~~ n
= 1,2,
$$
which are infrared dependent. To overcome this problem it is necessary to
introduce the infrared cut-off (or infrared regularization), to separate the
domain of large $k^2_{\bot}$, where perturbative QCD is reliable, from the
domain of small $k^2_{\bot}$. The contribution of the latter must be
addressed to noncalculable in perturbative QCD parton distribution. Such a
procedure is legitimate because of the factorization theorem which states
that the virtual photoabsorption cross section on the hadronic target $h,
\sigma^{\gamma}_h(x, Q^2)$ can be written down in the convolution form
%20
\be
\sigma^{\gamma}_h(x, Q^2) = \sum _{i} \sigma^{\gamma}_i(x, Q^2, M^2)\otimes
f_{i/h} (x, M^2),
\ee
where $\sigma^{\gamma}_i$ is the photoproduction cross section on the
$i^{th}$ parton $(i = q, \bar{q}, g), f_{i/h}$ are the parton distributions
in a hadron $h, \otimes$ stands for convolution. Both $\sigma^{\gamma}_i$
and $f_{i/h}$ depend on the infrared cut-off $M^2$, but the physical cross
section $\sigma^{\gamma}_hh$ is cut-off independent. The variation of $M^2$
corresponds to redistribution among partons: the trade of gluon for sea
quarks.

As follows from (19)
%21
\be
\Delta_g \Gamma_p = N_f \frac{< e^2 >}{2} \Delta g \overline{A(M^2)},
%22
\ee
where
\be
\overline{A(M^2)} = \int _{0} ^{1} dx_1 A(x_1, M^2)
\ee

The convenient way is to introduce cut off in $k^2 _{\bot}$
%23
\be
k^2_{\bot} > M^2 (x_1, p^2)
\ee
Generally, $M^2$ may depend on $x_1$ and $p^2$. For example, the cut-off in
quark virtuality in Fig.1 $-k^2 > M^2_0$ corresponds to the form (23) with
$M^2 = (1 - x_1)(M^2_0 + p^2 x_1)$ if $x_1 < - M^2_0/p^2$.

In the calculation of the diagrams of Fig.1 it is reasonable to neglect the
 light quark masses in comparison with gluon virtuality $p^2$, since we
expect that $\mid p^2 \mid$ is of order of characteristic hadronic masses,
$\mid p^2 \mid \sim 1 GeV^2 $~$^{[24]}$. Then, introducing the infrared cut-off
(23) we have $^{[35]}$
%24
\be
\overline{A(M^2)} = -\frac{\alpha_s}{2 \pi} \left \{1 -  \int \limits _0^1~
dx_1(1 - 2x_1) [lnr - r] \right \}
\ee
where
%25
\be
r = \frac{x_1(1 - x_1)p^2}{x_1(1 - x_1) p^2 - M^2 (x_1, p^2)} .
\ee
If $M^2 = Const$ -- a rectangular cut-off in $k^2_{\bot}$-- , the integral
in (24) vanishes (the integrand is antisymmetric under substitution $x_1
\rightarrow 1 - x_1$. Then $\bar{A} = -\alpha_s/2\pi$ and we obtain the
gluonic contribution to $\Gamma_{p,n}$ (12). However, other forms of
$M^2(x_1,p^2)$ result in different values of $\overline{A(M^2)}$, what
supports the claim $^{[25]}$ that $\overline{A(M^2)}$ is cut-off dependent.
Even more, if we put $p^2 = 0, m^2_q \not= 0$ -- the standard regularization
scheme in the calculation of nonpolarized deep inelastic scattering -- we
will find $\bar{A} = 0 ^{[33]}$. This result is, however, nonphysical
because the compensation of $-\alpha_s/2\pi$ term in $\bar{A}$ arises from
soft non-perturbative domain of $k^2_{\bot} \sim m^2_q$, which must be
attributed to sea quark distribution.

Although generally $\bar{A}$, as well as $\Delta g$ and $\Delta_g \Gamma_p$
are infrared cut-off dependent, a special class of preferable cut-off's can
be chosen. In OPE the mean asymmetry $\bar{A}$ is proportional to the
one-gluon matrix element of axial current
%26
\be
\Gamma_{\mu \lambda \sigma} (0, p, p) = < g, \epsilon_{\lambda} \mid j_{\mu
5} (0) \mid g,  \epsilon_{\sigma} >
\ee
at zero momentum transfer. It can be shown $^{[35]}$ that this quantity is
proportional to the divergence $l_{\mu} \Gamma_{\mu \lambda \sigma} (l, p_1,
p_2), ~ l = p_1 - p_2$ in the limit $l^2 \rightarrow 0$, i.e. to the one
gluon matrix element of axial anomaly.

The proof is the following. The general expression of the matrix element of
axial current over gluonic states with nonequal momenta has the form
$^{[35]}$
%27

$$
\Gamma_{\mu \lambda \sigma} (l, p_1, p_2) = F_1 (l^2, p^2)~ l_{\mu}
\epsilon_{\lambda \sigma \rho \tau} p_{1 \rho} p_{2 \tau} + $$
\be
\frac{1}{2}~ F_2 (l^2, p^2) \left [\epsilon_{\mu \lambda \sigma \rho} (p_1
+ p_2)_{\rho} + \frac{p_{1 \lambda}}{p^2} \epsilon_{\mu \sigma \rho \nu} p_{1
\rho} p_{2 \nu}-
 \frac{p_{2 \sigma}}{p^2} \epsilon_{\mu \lambda \rho \nu} p_{1 \rho}
p_{2 \nu} \right ]
\ee
where it was assumed that $p^2_1 = p^2_2 = p^2 < 0$. When deriving (27) we
used only Lorenz invariance, Bose symmetry of gluons and the conservation of
vector currents at gluonic vertices. The formfactors $F_{1,2}(l^2, p^2)$
have no kinematical singularities if $p^2 \not=0$ and, particularly,
$F_1(l^2, p^2)$ has no pole at $l^2 = 0$. From (27) we have:
%28

$$
l_{\mu} \Gamma_{\mu \lambda \sigma} = \left [F_2 (l^2, p^2) + l^2 F_1 (l^2,
p^2) \right ] \epsilon_{\lambda \sigma \rho \tau}~ p_{1 \rho}~p_{2 \tau}
$$
\be
\Gamma_{\mu \lambda \sigma} (0, p, p) = F_2 (0, p^2) \epsilon_{\mu \lambda
\sigma \rho} ~ p_{\rho}
\ee
and in the limit $l^2 \rightarrow 0$ the mentioned above statement follows.

Let us recall the derivation of the anomaly in perturbation theory
$^{[37]}$. The anomaly corresponds to the diagrams of Fig.2.

\vspace{2cm}
\hspace{10cm} Fig.2
\footnotesize

\hspace{10cm} The diagrams for the gluonic matrix

\hspace{10cm} element of axial current.

\vspace {2cm}

\normalsize

%\end{document}
These diagrams
have a superficial linear divergences. Cancellation of the divergences in
the diagrams Fig's 2a and 2b requires shifting of integration variable in
2b:
%nn
$$ k \rightarrow k - p_1 - p_2$$
This shift corresponds to substitution $x_1 \rightarrow 1 - x_1$ in the
integrated over $x_1$ diagrams of Fig.1, i.e. in eq.(24). We came to the
conclusion that in order to retain connection of $\Gamma_{\mu \lambda
\sigma}(0, p, p)$ with the anomaly which follows from (28) and to preserve
the standard form of the anomaly, the infrared cut-off $M^2(x_1, p^2)$ must
be symmetric under $x_1 \rightarrow 1 - x_1$. In this case $\bar{A} = -
\alpha_s/2 \pi$ and we come to the gluonic contribution to $\Gamma_{p,n}$
given by eq.12. With such a cut-off the anomaly comes entirely from the
hard momentum region and can be considered as a local probe of gluon
helicity. (It must be mentioned, however, that any other forms of
infrared cut-off are also possible, but they are less attractive). Some
care, however, is necessary, when the calculations with this cut-off are
compared with HO calculation in nonpolarized scattering, where as a
rule, another regularization procedure $-p^2 = 0, m^2_q \not= 0$ is used.

The gluonic contribution to the spin dependent proton structure function
$g_{1p}(x, Q^2)_{gl}$ can be found experimentally by measuring inclusive two
jets production in polarized DIS $^{[24]}$. However, only large
$-k^2_{\bot}$ component of Fig.1 diagrams can be determined in this way,
the small $-k^2_{\bot}$ component is nonmeasurable, since in this case it
is impossible to separate one jet from two jets events $^{[35]}$. This
fact is in complete accord with the formulated above statement about the
arbitrariness of infrared cut-off.

Now about the numerical values of the
constants $g_A$, $a_8$ and $\Sigma$ entering eq.(12). $g_A$ is known with a
very good accuracy, $g_A = 1.257 \pm 0.003$ $^{[5]}$.It must be mentioned,
however, that eq.(13$'$) follows from the assumption of exact isospin
symmetry. In fact, one may expect its violation of the order of 1\%.
Therefore, the uncertainty in this number is not given by experimental
error, but may be $\sim 1\%$.
%\end{document}
Under assumption of SU(3) flavour symmetry in baryon decays $a_8$ is equal
to
%29
\be
a_8 = 3F - D = 0.59 \pm 0.02
\ee
where $F$ and $D$ are axial coupling constants of baryon $\beta$-decays in
SU(3) symmetry and the numerical value in the r.h.s. of (19) follows
from the best fit to the data $^{[38]}$. The combination $3F-D$ can be found
also from any pair of baryon $\beta$-decays. The comparison of the values of
$3F-D$, obtained in this way shows, that the spread is rather narrow
$^{[39]}$, $\mid \delta a_8 \mid \leq 0.05$ and at least $a_8 > 0.50$. This
may be an argument in the favour that SU(3) violation is not large
here. (For the recent suggestion of such violation see,however,
$^{[40]}$).  It should be mentioned that at fixed $\Gamma_{p,n}$ the
uncertainty in $a_8$ only slightly influences the most interesting
quantities $\Sigma$ and $\Delta s$.  As follows from (12) and (14)
%30
\be
\delta \Sigma = - \delta (\Delta s) = \frac{1}{4} \delta a_8.
\ee
$a_8$ was
also determined by the QCD sum rule method $^{[41]}$. In this approach no
SU(3) flavour symmetry was assumed and the result
%31
\be
a_8 = 0.5 \pm 0.2
\ee
is in agreement with (29), although the error is large.

What can be said theoretically about $\Sigma$? In their famous paper
$^{[42]}$ Ellis and Jaffe assumed that the strange sea in the nucleon is
nonpolarized, $\Delta s = 0$. Then
%32
\be
\Sigma \approx a_8 \approx 0.60
\ee
(The sum rule (12) in the framework of this assumption is called
Ellis-Jaffe sum rule).This number is in a contradiction with the
experimental data pointing to smaller values of $\Sigma$. On the other hand,
Brodsky, Ellis and Karliner $^{[43]}$ had demonstrated that in the Skyrme
model at large number of colours $N_c$, $\Sigma \sim 1/N_c$ and is small.
{}From my point of view this argument is not very convincing: the Skyrme model
may be a good model for description of nucleon periphery, but not for the
internal part of the nucleon determining the value of $\Sigma$ (see also
$^{[44]}$).

\section{Calculations~ of~ the~ Matrix~ Elements~ over~ the
Polarized Nucleon by
QCD Sum Rule Approach}

%\end{document}
\vspace{2mm}
The QCD sum rule method was used to calculate the nucleon  coupling
constants $g_A ^{[45]}, a_8 ^{[41]}$ and the coefficients $b_{p,n}$ which
determine the twist-4 contributions $^{[13]}$. The basic features of the
approach are the same as originally suggested for calculation of nucleon
magnetic moments $^{[46, 47]}$.

We wish to find the diagonal matrix element
%33
\be
< p, s \mid~j (0)~ \mid p, s >
\ee
To this end add to the QCD Lagrangian the term
%34
\be
\Delta L = j(x)~S,
\ee
where $S$ is a constant external source. In the case of $g_A$
determination for $j$ and $S$ we substitute $j^{(3)}_{\mu 5}$ and
$A^{(3)}_{\mu}$ - the constant external axial field etc. Consider the
polarization  operator of the currents $\eta$ with the nucleon quantum
numbers
%35
\be
\Pi(p) = i ~\int~ d^4 x~ e^{ipx}~ < 0 ~ \mid~ T ~\left \{\eta(x), ~\bar{\eta}
(0)\right \} \mid 0~ >
\ee

%36
\be
\eta_p = \left (u^a(x) C \gamma_{\mu} u^b (x)\right ) \gamma_{\mu} \gamma_5
d^c (x) \epsilon^{abc}
\ee
$$ \eta_n = \eta_p(u \leftrightarrow d)$$
where
$a, b, c = 1,2,3$ are colour indeces. Separate in $\Pi(p)$ the term
proportional to external source $S$:
%37
\be
\Gamma(p,p,0)S = i^2 \int
d^4x d^4z e^{ipx} S< \mid T \left \{\eta(x), j(z), \bar{\eta}(0)\right \}
\mid 0 >
\ee
Suppose that $p^2 < 0, \mid p^2 \mid \gg R^{-2} _c$, where
$R_c$ is the confinement radius, and perform OPE in $1/p^2$. Unlike the
standard OPE for polarization operator, the presence of external source
results in appearance of a new type of vacuum expectation values (v.e.v.),
induced by external source. If the source $S$ is an axial field $A_{\mu}$,
then, e.g.
%38
\be
< 0 \mid \bar{u} \gamma_{\mu} \gamma_5 u \mid 0 >_A = c
A_{\mu}
\ee
It can be easily shown that for massless quarks $m_q = 0~~(q =
u, d, s), ~~~c = f^2_{\pi}$ in the case of isospin vector $A^{3}_{\mu}$ or
octet $A^8_{\mu}$ axial field. In the first case we can write
$$ < 0 \mid
\bar{u}\gamma_{\mu} \gamma_5 u \mid 0 >_A = \lim_{q \rightarrow 0}
\frac{1}{2} \int d^4 x e^{iqx} \times
$$
%39
\be
\times < 0 \mid T\left
\{j^{(3)}_{\nu5}(x), j^{(3)}_{\mu5} (0)\right \} \mid 0 > \equiv \lim_{q
\rightarrow 0} \Pi^{(3)}_{\nu \mu}(q) A^{(3)}_{\nu}
\ee
Since the axial
current $j^{(3)}_{\mu}$ conserves
%40
\be
\Pi^{(3)}_{\nu \mu}(q) =
-(\delta_{\mu \nu} q^2 - q_{\mu} q_{\nu}) \Pi (q^2)
\ee
The non-zero result
for $\Pi_{\mu \nu}(q)$ at $q \rightarrow 0$ comes from the pole $\sim 1/q^2$
in $\Pi(q^2)$ which corresponds to the massless pion intermediate state
%41
\be
\Pi^{(3)}_{\mu \nu}(q)= \left (-\frac{q_{\mu} q_{\nu}}{q^2 - \mu^2_{\pi}}+
\delta_{\mu \nu}\right )f^2_{\pi}
\ee
In (41) the pion mass in the
propagator was accounted as the most important effect of nonvanishing quark
masses at small $q^2$. Substituting (41) into (39) and going to the limit $q
\rightarrow 0$ we obtain the desired result.

The same proof may be repeated for $A^{(8)}_{\mu}$ in the limit of $SU(3)$
flavour symmetry. The vacuum expectation values which enter the OPE of the
r.h.s. of (36) can be classified according to their dimensions and the
number of loops  (the dimension of $A_{\mu} = A^{(3)}_{\mu}, A^{(8)}$ is
1, the chirality conserving structure is considered)

\vspace{3mm}
%42
\be
\begin{tabular}{ccc}
Dimension &           V.e.v.                   & Number of loops\\
\\
    1     & $1 .  A_{\mu}$                    &         2       \\
\\
    3     & $< 0 \mid\bar{q} \gamma_{\mu}\gamma_5 q\mid 0 >_A = f^2_{\pi}
A_{\mu}$                                        &         1        \\
\\
    5     & $< 0 \mid \alpha_s G^2_{\mu \nu} \mid 0 > A_{\mu}$& 2    \\
\\
    5     & $< 0 \mid \bar{q} \gamma_{\nu} \tilde{G}^n_{\nu \mu}
\frac{1}{2} \lambda^n q \mid 0 >_A = f^2_{\pi} m^2_1 A_{\mu}$ &        1 \\
\\
    7     & $< 0 \mid \bar{q} q \mid 0 >^2 A_{\mu}$&        0  \\
\end{tabular}
\ee

\vspace{3mm}
Here $\tilde{G}_{\mu \nu} = (1/2) \epsilon_{\mu \nu \lambda \sigma}
G_{\lambda \sigma},~~ m^2_1 \approx 0.2~ GeV^2$ $^{[48]}$. The important
role of induced by external field v.e.v. is evident -- besides the unit
operator they correspond to lowest dimensions and to minimal number of
loops. The OPE for the r.h.s. of (37) can be constructed exploiting the
v.e.v. given in (42).

In order to find the matrix element (33), represent the l.h.s. of (37) in
terms of contributions of physical states using dispersion relations.
Generally, when the momentum of external field is non-zero, $\Gamma$ is a
function of three variables, $\Gamma = \Gamma(p^2, p^2_2; q^2)
$, and may be
represented by the double dispersion relation:
%43
\be
\Gamma(p^2_1, p^2_2; q^2) = \int \limits^{\infty}_{0}~ ds_1~ \int
\limits^{\infty}_{0}~ ds_2 ~\frac{\rho(s_1, s_2; q^2)}{(s_1 - p^2_1)(s_2 -
p^2_2)} + P(p^2_1) f(p^2_2, q^2)+
\ee
$$+ P(p^2_2)~ f(p^2_1, q^2)$$,
where $P(p^2)$ is a polynomial and $f(p^2, q^2)$is given by the ordinary
dispersion relation
%44
\be
f(p^2, q^2) = \int \limits^{\infty}_{0} ~
\frac{\varphi(s, q^2)}{s - p^2} + subtr.terms
\ee
We are interested in the
limit $q \rightarrow 0, p^2_1 = p^2_2 = p^2$ and at the first sight it seems
that one vartiable dispersion relation for $\Gamma(p^2, p^2, 0)$ can be
written in this case. Indeed,
%45
\be
\int \limits^{\infty}_{0}~ ds_1~ \int
\limits^{\infty}_{0}~ ds_2 ~\frac{\rho(s_1, s_2; 0)}{(s_1 - p^2)(s_2 -
p^2)} = \int \frac {\rho (s_1, s_2, 0)}{s_1 - s_2} ds_1 ds_2 \left
(\frac{1}{s_2 - p^2} - \frac{1}{s_1 - p^2}\right )
\ee
In the first (second) term in the r.h.s. of (45) integration over
$s_1(s_2)$ can be performed and the result has the form of one-variable
dispersion relation. Such transformation is, however, misleading, because,
in general, the integrals
%nn
$$
\int ds_1 \frac{\rho(s_1, s_2, 0)}{s_1 - s_2}= -\int~ds_2 \frac{\rho(s_1, s_2,
0)}{s_1 - s_2}
$$
are ultraviolate divergent. This ultraviolate divergence cannot be cured by
subtractions in one-variable dispersion relation: only subtractions in the
double dispersion representation (43) can be used. It is evident that the
procedures which kill the subtraction terms and lead to fast convergence of
dispersion integrals in standard one-variable dispersion representations
like Borel transformation in $p^2$ do not help here.

Let us duscuss the determination of the coupling constant $a_8$ in more
details. Consider in (37) the coefficient function at the structure
$(A^{(8)}_{\mu} p_{\mu}) \hat{p} \gamma_5$. The advantage of this
structure is that it has the largest power of momenta in the numerator.
Therefore, the integrals for the coefficient function converge better and
the contribution of excited states in the dispersion relation (43), which
is the background in our calculation, is smaller. It can be easily seen
that the function $\rho$ in the r.h.s. of (43) which correspond to the
bare loop diagram Fig.3 has in this case the form
%46
\be
\rho(s_1, s_2) = a s_1 s_2 \delta(s_1 - s_2),
\ee
where $a$ is a calculable constant

\vspace{3cm}
\footnotesize
\begin{center}
Fig.3. The bare loop diagram, corresponding to determination of the
coupling constants $g_A$ or $a_8$. The solid lines correspond to quark
propagators, crosses mean the action of currents $\eta, \bar{\eta}$, the
bubble corresponds to quark interaction with external field.
\end{center}

\vspace {3mm}
\normalsize

The substitution of (46) into (43) gives for the first term in the OPE in
the r.h.s. of (37) at $p^2_1 = p^2_2 \equiv p^2$
%47
\be
\Gamma(p^2) = a~ \int ^{\infty} _{0}~ \frac{s^2_1 ds_1}{(s_1 - p^2)^2}
\ee
%\end{document}
In this simple example the dispersion representation is reduced to
one-variable dispersion relation, but with the square of $(s_1 - p^2)$ in
the denominator. Of course, by integrating by parts (47) may be transformed
to the standard dispersion representation. However, the boundary term
arising at such transformation must be accounted; it does not vanish even
after application of the Borel transformation. This means that even in this
simplest case the representation (43) is not equivalent to one-variable
dispersion relation.
\newpage

\vspace*{5cm}
\footnotesize
\begin{center}
Fig.4. The integration domains the in $s_1, s_2$ plane
\end{center}
\vspace{5mm}
\normalsize
Let us represent $\Gamma(p^2, p^2; 0)$ in terms of contributions of
hadronic states using (43) and separating the contribution of the lowest
hadronic state in the channels with momentum $p$ $^{[49]}$. As is seen from
Fig.4, it is convenient to divide the whole integration region in $s_1, s_2$
into three domains: I) $0 < s_1 < W^2,~~ 0 < s_2 < W^2$; ~ II) $0 < s_1 <
W^2,~ W^2 < s_2 < \infty; ~~~ W^2 < s_1 < \infty, ~~~~~ 0 < s_2 < W^2;$~
III)$W^2 < s_1 < \infty, ~ W^2 < s_2 < \infty$. Adopt the standard in QCD
sum rule model of hadronic spectrum: the lowest hadronic state nucleon plus
continuum, starting from some threshold $W^2$. Then in the domain I only the
lowest hadronic state $N$ contributes and
%48
\be
\rho(s_1, s_2) = a_8
\lambda^2 \delta (s_1- m^2) \delta (s_2 - m^2)
\ee
where $m$ is the nucleon mass
 $\lambda$ is the transition constant of the nucleon in the
current $\eta$: \\
$< N \mid \bar{\eta} \mid 0 > = \lambda
\bar{\upsilon}$ where $\upsilon$ is the nucleon spinor. In the domain III
the higher order terms in OPE may be neglected and the contribution of
hadronic states is with a good accuracy equal to the contribution of the
bare quark loop (like Fig.3) with perturbative corrections. The further
application of the Borel transformation in $p^2$ essentially suppresses this
contribution.

The consideration of the domain II contribution is the most troublesome and
requires an additional hypothesis. Assume, using the duality arguments, that
in this domain also, the contribution of hadronic states is approximately
equal to the contribution of the bare quark loop. The accuracy of this
approximation may be improved by subtraction from each strip of the domain
II of the lowest hadronic state contributions proportional to $\delta(s_1 -
m^2)$ or $\delta(s_2 - m^2)$. The terms of the latter type also persist in
the functions $f(p^2_1), f(p^2_2)$ in (43). They correspond to the process
when the current $\bar{\eta}$ produces the nucleon $N$ from the vacuum and
under the action of the external current $j$ the transition to excited state
$N \rightarrow N^{\ast}$ occurs or vice versa (Fig.5).

\vspace{2.5cm}

\footnotesize
\begin{center}
Fig.5. The schematical representation of $N \rightarrow N^{\ast}$ $(N^{\ast}
\rightarrow N)$ transitions in the external field.
\end{center}

\vspace{5mm}

\normalsize
At $p^2_1 = p^2_2 = p^2$ these contributions have the form
%49
\be
\int ^{\infty} _{W^2} \frac {b(s)ds}{p^2 - m^2) (s - p^2)}
\ee
with some unknown function $b(s)$. The term (49) will be accounted
separately in the l.h.s. of (43). I stress that the term (49) must be
added to the l.h.s. of (43) independently of the form of the bare loop
contribution $\rho(s_1, s_2)$. Even if $\rho(s_1, s_2) = 0$, when the OPE
for the vertex function $\Gamma(p^2, p^2, 0)$ with zero momentum transfer
starts from condensate terms - the term (43) may persist. (49) may be
written as
%50
\be
\int ^{\infty} _{W^2}~ dsb(s) \left (\frac{1}{p^2 - m^2} + \frac{1}{s - p^2}
\right )~ \frac{1}{s - m^2}
\ee
The functions $f(p^2)$ in (43) can be represented by dispersion relation as
%51
\be
f(p^2) = \int ^{\infty} _0 \frac {d(s)} {s - p^2} ds
\ee
 The integration domain (51) may be also divided into parts $0 < s < W^2$
 and $W^2 < s < \infty$.\\
 According to our model the contribution of the
 first part is approximated by $N$-state contribution, the second one by
 continuum. These two parts look like the contributions of the first and the
 second terms in the bracket in (50).

Now we can formulate the recipe how the sum rule can be written. At the
phenomenological side -- the l.h.s. of the sum rule -- there is a
contribution of the lowest hadronic state $N$ and the unknown term (50),
corresponding to nondiagonal transition $N \rightarrow N^{\ast}$ in the
presence of external field;
%52
\be
\frac {\lambda^2 a_8}{(p^2 - m^2)^2} + \int ^{\infty}_{W^2}~ dsb(s)
\frac{1}{s - m^2}\left (\frac{1}{p^2 - m^2} + \frac {\alpha(s)}{s - p^2}
\right )
\ee
The contribution of continuum corresponding to the bare loop (or also to the
higher order terms in OPE, if their discontinuity does not vanish at $s
\rightarrow \infty$) is transferred to the r.h.s. o the sum rule. Here it is
cancelled by the bare loop contribution from the same domain of integration.
As a result, in the double dispersion representation of the bare loop the
domain of integration over $s_1, s_2$ is restricted to $0 < s_1, s_2 < W^2$.
Finally, apply the Borel transformation in $p^2$ to both sides of the sum
rule. In the r.h.s. -- QCD side -- the contribution of the bare loop has the
form
%53
$$
\int ^{W^2} _{0}~ ds_1~ \int ^{W^2} _{0}~ ds_2 \rho(s_1, s_2) \frac{1}{s_1 -
s_2} [e^{-s_2/M^2} - e^{-s_1/M^2}]
$$
\be
= 2P~ \int^{W^2} _{0} ds_2 \int^{W^2} _{0}~ ds_1 \frac{\rho(s_1, s_2)}{s_1 -
s_2} e^{-s_2/M^2}
\ee
where $P$ means the principal value and the symmetry of $\rho(s_1, s_2)$ was
used. The l.h.s. of the sum rule is equal to
%54
\be
a_8 \frac{\lambda^2}{M^2} e^{-m^2/M^2} - Ae^{-m^2/M^2} + e^{-m^2/M^2}
{}~\int^{\infty}_{W^2}~ dsb(s) \frac{\alpha(s)}{s - m^2} exp [ - (s -
m^2)/M^2]
\ee
where
%55
\be
A = \int ^{\infty}_{W^2}~ ds \frac{b(s)}{s - m^2}
\ee
In (54) $A$  is an unknown constant which can be determined from the same
sum rule exploiting the fact that the $M^2$ dependence of the second term
in (54) differs from the first. Of course, the calculation is reliable only
if the contribution of the second term in (54) is smaller, say, less than
30\% comparing with the first. The last term in (54) is of the same origin
as the second one -- it comes from inelastic transitions $N \rightarrow
N^{\ast}$ (Fig.5). but it is exponentially suppressed in comparing with the
second and as a rule may be neglected.

Omitting the details of the calculation, I present the result -- the sum
rule for determination of the coupling constant $a_8$ $^{[41]}$
%56
$$
\tilde{\lambda}^2_N \left (\frac{a_8}{M^2} + A \right ) e^{-m^2/M^2} = -M^4
E_2 \left (\frac{W^2}{M^2} \right ) L^{-4/9}-- \frac{1}{4}b E_0 \left
(\frac{W^2}{M^2}\right ) L^{-4/9}
$$
\be
+\frac{a^2}{M^2} \left (- \frac{4}{3} + \frac{8}{9} \right ) L^{4/9} +
\frac{16}{3} \pi^2 f^2_{\pi} M^4 E_1 \left (\frac{W^2}{M^2} \right )L^{-4/9}
+ \frac{28}{9} (2 \pi)^2 f^2_{\pi} m^2_1 E_0 \left (\frac{W^2}{M^2}\right )
L^{-8/9}
\ee
Here
%57
$$
a = -(2 \pi)^2 < 0 \mid \bar{q} q \mid 0 > = 0.55 GeV^3, ~ b = (2\pi)^2 < 0
 \mid \frac{\alpha_s}{\pi} G^2 \mid 0 > =  0.5 GeV^4$$
\be
 E_0(x) = 1 - e^{-x},
{}~~ E_1(x) = 1 - (1 + x)e^{-x}, ~~ E_2(x) = 1 - \left (1 + x +
\frac{x^2}{2}\right )e^{-x}
\ee
%58
\be
L = ln \left (\frac{M}{\Lambda}\right ) /
ln \left (\frac {M}{\Lambda} \right )
\ee
For the nucleon coupling constant
$\tilde{\lambda}^2_N =32 \pi^4 \lambda^2$ and continuum threshold we took
the values found in the mass sum rules $^{[50, 46]}$, $\tilde{\lambda}^2_N
= 2.1 GeV^6, W^2 = 2.3 GeV^2$. The normalization point $\mu$ was chosen as
$\mu = 0.5 GeV$. The sum rule (56) may be compared with the sum rule
determining the nucleon mass $^{[50, 46]}$:
%59
\be
\tilde{\lambda}^2_N \frac{1}{M^2}
e^{-m^2/M^2} = M^4 E_2 \left (\frac{W^2}{M^2} \right ) L^{-4/9} +
\frac{1}{4} < 0 \mid \frac{\alpha_s}{\pi} G^2 \mid 0 > E_0 \left
(\frac{W^2}{M^2} \right ) L^{-4/9} + \frac{4}{3} \frac{a^2}{M^2} L^{4/9}
\ee
Adding (56), (59) and applying to the sum the differential operator
$$ (1 -
M^2 \frac {\partial}{\partial M^2}) M^2 e^{-m^2/M^2} $$
which kills the constant
$A$, we get the sum rule for $a_8$:
%60
$$ a_8 = -1 + \frac{8}{9}
\frac{1}{\tilde{\lambda}^2_N} \left (1 - M^2 \frac{\partial}{\partial M^2}
\right ) e^{m^2/M^2} \left \{6 \pi^2 f^2_{\pi} M^4 E_1 \left
(\frac{W^2}{M^2} \right ) L^{-4/9} + \right.$$ \be \left.+ 14 \pi^2
f^2_{\pi} m^2_1 M^2 E_0 \left (\frac{W^2}{M^2} \right ) L^{-8/9} + a^2
L^{4/9} \right \}
\ee
The $M^2$ dependence of the r.h.s. of (60) is plotted
in Fig.6.

\vspace{4cm}
\footnotesize
\begin{center}

Fig.6.

The Borel parameter $M^2$ dependence of the r.h.s. of (60).
\end{center}

\vspace{5mm}

\normalsize
It is seen that
the $M^2$ dependence in the interval $0.8 < M^2 < 1.3 GeV^2$, where the
nucleon mass sum rule holds, is remarkable.  Also the constant $A$ found
from (56) is not small enough, $A \simeq -0.3$.  For these reasons the
error in the determination of $a_8$ is large and the final result is
%61
\be
a_8 = 0.5^{+0.25}_{-0.15}
\ee
I would like to mention that the uncertainty
in the value of $a_8$ arising from proportional to $A$ term in the l.h.s. of
(56) manifests itself as well in the lattice calculations of these or
similar quantities. In the lattice calculations the polarization operator
(35) is measured at large euclidean times $t$ and the result is proportionl
to
%62
\be \tilde{\lambda}^2_n~e^{-mt}~[a_8 (1 + mt) + 2 Am^2 ]
\ee
(see the
second reference in $^{[46]}$ where a similar formula was obtained in the
case of nucleon magnetic moments). The last term is usually neglected in
lattice calculations. As is seen from (62) and from the values of $a_8$ and
$A$ presented above, in order to find $a_8$ with 10\% accuracy it is
necessary to go to $t > 10 m^{-1} \approx 2fm$, where the whole effect is
very small.

I dwell now on the attempt to determine the proton singlet axial coupling
constant $a_0 =\Sigma$ by QCD sum rules $^{[51]}$. The difference in
comparison with $g_A$ and $a_8$ determination comes from nonconservation of
singlet axial current caused by the anomaly
%63
\be
\partial_{\mu} j^{(0)}_{\mu 5} = \frac{3 \alpha_s}{4 \pi}~ G^n_{\mu \nu}~
\tilde{G}^n_{\mu \nu} + 2 im_s~ \bar{s} \gamma_5 s
\ee
In the r.h.s. the strange quark mass term is accounted. whose contribution
will be essential. The flavour singlet pseudoscalar meson $(\eta')$ is not a
Goldstone and the used above method (eqs.(38)-(41)) of determination of
induced by the axial field v.e.v.'s fails. So, a special investigation is
needed in order to find  the value of quark condensate induced by axial
field. Like (39) we can put
%64
\be
< 0 \mid \bar{u} \gamma_{\mu} \gamma_5 u \mid 0 >_A = lim_{q \rightarrow 0}~
\Pi^{(0)}_{\nu \mu} (q) A^{(0)}_{\nu}
\ee
where $\Pi^{(0)}_{\nu \mu}$ is defined by eq.39 with substituting
$j^{(3)}_{\mu 5} \rightarrow (2/3) j^{(0)}_{\mu 5}$. But now
$\Pi^{(0)}_{\nu \mu}$ contains both -- transverse and longitudinal parts
%65
\be
\Pi^{(0)}_{\nu \mu} (q) = -\Pi_L (q^2) \delta_{\mu \nu} - \Pi_t(q^2)
(\delta_{\mu \nu} q^2 - q_{\mu} q_{\nu}
\ee
$\Pi_L (q^2)$ and $\Pi_t(q^2)$ have no poles at $q^2 =0$ and the interesting
 for us quantity is
%66
\be
f^2_0 = -\tilde{\Pi}_L(0) = [\Pi_L(0) - \Pi_{L, pert}(0) ]
\ee
The perturbative part was subtracted in (66) since it will be accounted in
explicit way by perturbative calcullation (practically, it is small). The
constant $f^2_0$ plays here the same role as $f^2_{\pi}$ in (38).

In order to separate $\Pi_L(q^2)$ multiply $\Pi_{\mu \nu}$ by $q_{\mu}
q_{\nu}$. We have
%67
$$
q_{\mu} q_{\nu} \Pi^{(0)}_{\mu \nu} (q) = -q^2 \Pi_L(q^2) = i~
\frac{\alpha_s}{4 \pi} \int~ d^4x~ e^{iqx} < 0 \mid T \left \{G^n_{\mu \nu}
(x) \tilde{G}^n_{\mu \nu}(x), \right.$$
\be
\left.\frac{3 \alpha_s}{4 \pi}~
G^n_{\alpha \beta}(0) \tilde{G}^n_{\alpha \beta}(0) + 2 im_s~ \bar{s}
\gamma_5 s(x) \right \}\mid 0 >
\ee
Represent $\Pi^{(0)}_{\mu \nu}(q)$ in terms of contributions of physical
states using the dispersion relations. Then it can be easily seen that
$\Pi_L(q^2)$ is contributed by pseudoscalars, the axial mesons contribute to
$\Pi_t(q^2)$. Using the notations
%68
\be
< 0 \mid \bar{q} \gamma_{\nu} \gamma_5 q \mid \eta' > = ig^q _{\eta'}~
q_{\nu} ~~~ <0 \mid j^{(0)}_{\nu 5} \mid \eta' > = i f_{\eta'} q_{\nu}
\ee
$(q = u,d)$ and representing $\Pi_L(q^2)$ in terms of $\eta'$ contribution
 and continuum, we write
%69
\be
\Pi_L (q^2) = \frac{g^u_{\eta'}~f_{\eta'}~m^2_{\eta'}}{m^2_{\eta'} - q^2} +
\frac{1}{\pi}~ \int \limits_{s_0}^{\infty}~ \frac{\beta(s')}{s' - q^2}~ ds'
\ee
where $s_0$ is the continuum threshold. In (69) $\beta(s)$ is determined by
perturbative calculation of the bare loop (Fig.7) corresponding to the r.h.s
of (67).

\vspace{2.5cm}
\footnotesize
\begin{center}
Fig.7. The bare loop contribution to the r.h.s. of (67),

the dashed lines are gluons.
\end{center}
\vspace{5mm}

\normalsize
Therefore, $\tilde{\Pi}_L(q^2)$ is given by
%70
\be
\tilde{\Pi}_L(q^2) \equiv \Pi_L(q^2) - \Pi_{L~pert}(q^2) =
\frac{g^u_{\eta'}~f_{\eta'}~m^2_{\eta'}}{m^2_{\eta'} - q^2} - \int \limits
_{0}^{s_0}~\frac{\beta(s')}{s' - q^2}~ ds'
\ee

%\newpage

To find the quantity $\tilde{\Pi}_L(0)$ consider the integral over the
contour $C$ in Fig.8 in the complex plane $q^2$:

\vspace{3.5cm}

\footnotesize
\begin{center}
Fig.8. The integration contour in the complex plane in eq.(71).
\end{center}
\vspace{5mm}

\normalsize
%71
\be
\frac{1}{2 \pi i}~\int \limits _{C}~ \tilde{\Pi}_L (q^2)~ \frac{dq^2}{q^2} =
\tilde{\Pi}_L(0) = - g^u_{\eta'} f_{\eta'} +  \frac{1}{\pi} \int \limits
_{0}^{s_0}~ \beta(s) \frac{ds}{s}
\ee
The calculation of the diagram Fig.7 gives
%72
\be
\beta(s) = \frac{3~\alpha^2_s}{8 \pi^3}~s
\ee
To find $ g^u_{\eta'} f_{\eta'}$ let us write OPE for the r.h.s. of eq.(67)
 and use the QCD sum rule method. In OPE take into account, besides the bare
 loop, the contribution of the gluon condensate and of the quark-gluon
 condensate (Fig.9)
%73
\be
-g < 0 \mid \bar{s} \sigma_{\alpha \beta} \frac{1}{2} \lambda^n~G^n_{\alpha
\beta}~s \mid 0 > \equiv m^2_0 < 0 \mid \bar{s} s \mid 0 >,
\ee
$m^2_0 = 0.8 GeV^2$, $^{[50]}$ stemming from the term, proportional to $m_s$
in (67).

%\newpage
\vspace{2.5cm}
\footnotesize
\begin{center}
Fig.9. Gluon~(a) and quark-gluon~(b) condensate contributions to the r.h.s.
of eq.(67).
\end{center}
\normalsize

\vspace{5mm}
Let us also account the condensate $\sim < 0 \mid g^3 G^3 \mid 0 >$, taking
its estimate from Ref.52.

The sum rule for $g^u_{\eta'} f_{\eta'}$ is given by:
%74
$$
g_{\eta'} f_{\eta'} = \frac{1}{m^2_{\eta'}}~e^{m^2_{\eta'}/M^2} \frac{3
\alpha_s(M^2)}{16 \pi^3} [\frac{2\alpha_s(M^2)}{\pi}~ M^4 E_1\left (\frac
{s_0}{M^2}\right )+$$
\be
+ \left (1 + \frac{\epsilon}{M^2}\right ) + \frac{4}{3}~m^2_0 m_s a_s
\frac{1}{M^2}
\ee
Here $\epsilon \approx 0.2 GeV^2$ $^{[52]}$
%75
\be
a_s = -(2\pi)^2 < 0 \mid \bar{s} s \mid 0 > \approx 0.8 a \approx 0.44 GeV^3
\ee
{}From (74) it numerically follows
%76
\be
g^u_{\eta'}~ f_{\eta'} = (4 \pm 1)\cdot 10^{-3}~ GeV^2
\ee
and
%77
\be
f^2_0 = -\tilde{\Pi}_L(0) = g^u_{\eta'}~f_{\eta'} -
\frac{3\alpha_s}{8\pi^3}~s_0 \approx 3.5.10^{-3} GeV^2
\ee
in comparison with the similar constant for the octet current $f^2_{\pi} =
(0.133~ GeV)^2 = 18.10^{-3}~ GeV^2$.

Nevertheless that all looks okay in this calculation, the result is wrong.
In order to demonstrate this, determine the $s$-quark coupling constant
$g^s_{\eta'}~f_{\eta'}$. The only diference with the previous calculation
will be that in the r.h.s. of eq.(67) the equal-time commutator,
proportional to $m_s \bar{s}s$ will appear and, as a consequence, owing to
the $s$-quark condensate an additional term appears in the r.h.s. of the sum
rule (74)
%78
\be
-\frac{1}{m^2_{\eta'}}~e^{m^2_{\eta'}/M^2} \frac{4}{3} m_s < 0 \mid \bar{s}
s \mid 0 > \approx 6.10^{-3}~ GeV^2
\ee
Thus, our calculation results in a
wrong ratio of $\eta'$ interaction constants with $u$ (or $d$) and
$s$-quarks
%79
\be
\frac{g^s_{\eta'}}{g^u_{\eta'}} \approx 2.5
\ee
whereas
it is known that $\eta'$ is a flavour singlet and this ratio must be close
to 1.

We arrive at the conclusion that the sum rule for the longitudinal
polarization operator with the singlet axial current does not work at the
values of the Borel parameter $M^2$ of order of the $\eta'$ mass square. The
only way to avoid this discrepancy is to assume that in the OPE in this case
there are some important higher order terms. This conclusion is not
surprising:it has been known long ago $^{[53,54]}$ that with the standard
OPE it is impossible to describe the Okubo-Zweig-Iizuka rule violation in
the pseudoscalar and longitudinal axial channels. It is necessary to take
into account higher order terms and instantons in the direct channel were
proposed as possible candidates for such terms. Now this idea is strongly
supported by the calculations in the instanton liquid model $^{[55]}$, as
well as by the lattice calculations $^{[56]}$.

Since the attempt to calculate the induced by singlet axial vacuum
condensate fails, it is impossible to find by QCD sum rules the coupling
constant $a_0 = \Sigma$ -- the part of the proton spin projection carried by
quarks. In fact, the situation is even wose. It can be shown $^{[51]}$ that
if the value of $\Sigma$ is taken from experiment, then the sum rule or its
determination (like (60) with the induced by the field v.e.v. considered as
free parameters) and the sum rule for transverse singlet axial current
polarization operator $\Pi_t(q^2)$ are in contradiction . This  indicates the
OPE breaking down at virtualities $\sim 1GeV^2$ in the vertex function for
proton interaction with singlet axial current or/and in the transverse
singlet  axial current polarization operator. Thus, the situation in the
singlet axial channel resembles the one in the pseudoscalar channel and one
may expect a noticeable violation of the OZI rule here too. A similar
trouble, perhaps, faces attempts to determine $\Sigma$ using the so called
"$U(1)$ Goldberger-Treiman relation" (for a review see $^{[57]}$). So, at
this stage, the only way to find $\Sigma$ is from experiment, exploiting
eq.12.

\vspace{5mm}
\section{Twist-4 Corrections to $\Gamma_{p,n}$ from QCD Sum Rules}

\vspace{3mm}

The general theory of twist 4 corrections to deep inelastic scattering on
 polarized nucleons has been developed by Shuryak and Vainstein $^{[58]}$.
 The have found (see Ref.13,Errata for corrections of errors)
 %80
\be
(\Gamma_p \pm \Gamma_n)_{twist 4} = - \frac{8}{9}~ \frac{C^{S,NS}}{Q^2}~
\left [~\langle \langle U^{S,NS} \rangle \rangle - \frac {m^2}{4}~ \langle
\langle V^{S,NS} \rangle \rangle \right ]
\ee
$$
+ \frac{2}{9}~
\frac{m^2}{Q^2}~ \int \limits_{0}^{1}~ dx~x^2 g_{1, p \pm n} (x) $$
Here indeces $S, NS$ correspond to + and - signs in the l.h.s. of (80), $C^S
= 5/18, C^{NS} = 1/6$. The reduced matrix elements $\langle \langle U
\rangle \rangle, \langle \langle V \rangle \rangle$ are related to the
matrix elements of the operators
%81
\be
U^u_{\mu} = \bar{u} g ~\tilde{G}^n_{\mu \nu} \gamma_{\nu} \frac{1}{2}
\lambda^n u
\ee
%82
\be
V^u_{\mu \nu, \sigma} = \frac{1}{2}~\bar{u} g~ \tilde{G}^n_{\mu \nu}
 \gamma_{\sigma} \frac{1}{2} \lambda^n u + (\nu \rightarrow \sigma)
\ee
in the following way
%83
\be
\langle N \mid U_{\mu} \mid N \rangle = s_{\mu} \langle \langle U \rangle
\rangle
\ee
%84
\be
\langle N \mid V_{\mu \nu, \sigma} \mid N \rangle = S_{\nu, \sigma} A_{\mu,
\nu}~s_{\mu} p_{\nu} p_{\sigma} \langle \langle V \rangle \rangle
\ee
where $s_{\mu}$ is the unit nucleon spin vector, $A_{\mu, \nu}$ and $S_{\nu,
\sigma}$ stand for (anti)symmetrization over the given subscripts. The
indeces $S, NS$ mean
%85
\be
S \rightarrow \bar{u} u + \bar{d} d +\frac{18}{5} \bar{s} s, ~~~ NS
\rightarrow \bar{u} u - \bar{d} d
\ee
$\langle \langle U \rangle \rangle, \langle \langle V \rangle \rangle$ were
calculated by Balitsky, Braun and Kolesnichenko (BBK) $^{[13]}$ using the
QCD sum rule approach. The result for $b_{p-n}$ was given in (10). For
$b_{p+n}$ it was obtained
%86
\be
b_{p+n} = -0.022~ GeV^2
\ee
This result, however, cannot be considered as reliable for the following
reasons:

1.BBK use the same hypothesis as Ellis and Jaffe did, i.e., assume that
$s$-quarks do not contribute to the spin structue functions and instead of
singlet (in flavour) operator consider the octet one.

2. When determining the induced by external field vacuum condensates, which
are very impotant in such calculations they saturate the
corresponding sum rule by $\eta$ meson contribution, what is wrong. (Even
the saturation by $\eta'$-meson would not be correct since $\eta'$ is not a
Goldstone).

3. One may expect that in the same way as in the calculation of $\Sigma$ by
the QCD sum rule, in this problem the OPE series diverge at the scale $\sim
1~ GeV$ where the BBK calculation proceeds.

Even in the case of the Bjorken sum rule, where the mentioned above problems
are absent, the value $b_{p-n}$ obtained by BBK is questionable. Let us
consider this calculation in more details. The bare loop diagram for this
case is shown in Fig.10.

\vspace{3cm}
\footnotesize
\begin{center}
Fig.10. The bare loop diagram for twist 4 correction to the Bjorken sum rule

for deep inelastic electron-nucleon scattering. The dashed line corresponds

to discontinuity over $p^2_1$ at $p^2_1 \not= p^2_2$.
\end{center}

\vspace{5mm}

\normalsize
In $^{[13]}$ this diagram was calculated by introducing an ultraviolet cut
off $\mu^2$. It was found that for the chosen Lorentz structure the singular
in $p^2$ term is proportional to $p^4~ln^2(\mu^2- p^2)$. Such cut off
dependence reflects the fact that the spectral function $\rho(s_1, s_2)$ in
eq.(43) is not proportional to $\delta$-function. The logarithm square
dependence of $\Gamma(p^2)$ on the cut-off cannot be removed by Borel
transformation  .For these reasons, in order to obtain physical results the
authors of ref. $^{[13]}$ considered various values of $\mu^2$ in the
interval $0.1< \mu^2 < 1~ GeV^2$ and included uncertainties arising from this
procedure into the error. From the presented above point of view such an
approach is not legitimate. In this case $\rho(s_1, s_2)$ is proportional to
$s_1, s_2$:
$$
\rho(s_1, s_2)= bs_1 s_2
$$
where $b$ is a constant. In the model of hadronic spectrum accepted in
Sec.2, we have after the Borel transformation and using eq.(53)

\newpage
$$
\Gamma(M^2) = 2P~ \int _{0}^{W^2}~~ ds_2~\int_{0}^{W}~~ ds_1
\frac{\rho(s_1, s_2)}{s_1 - s_2} e^{s_2/M^2}$$
%87
\be
= 2b~ \int _{0}^{W^2}~~sdse^{-s/M^2} \left [W^2 + ln~ \frac{W^2}{s}\right ]
\ee
Eq.(27) essentially differs from the corresponding expression for the bare
loop contribution in ref.$^{[13]}$: e.g., the integrand in (87) is positive,
while in $^{[13]}$ it is negative in the main region of integration. Of
course, the QCD sum rule calculation in this case has a serious drawback: the
continuum threshold $W^2$ dependence of the result is not in the form of a
small correction of the type $exp(-W^2/M^2) \ll 1$ at $W^2 \gg M^2$, but
much more strong. This is a direct consequence of high (equal to 5) dimension
of the operators $U_{\mu}, V_{\mu \nu, \sigma}$. It is clear that the
higher is the dimension of the considered operator, the stronger will be
dependence on the continuum threshold and less certain the results of the QCD
sum rules calculations. It must be emphasized that for operators of high
dimensions the loop diagrams are in principle nonrenormalizable, the role of
excited states in the physical spectrum increases and the determination of
the lowest state contribution becomes impossible.

Eq.(87) must be taken instead of the contribution of the bare loop diagram,
used in $^{[13]}$. A similar procedure must be also applied in the case of
other terms in the sum rules $^{[13]}$ containing an ultraviolet cut-off.

BBK accounted for operators up to dimension 8. With the corrections
described above, the sum rules found by BBK have the form
%88
$$
\langle \langle U^{NS} \rangle \rangle + A^{NS}_U~M^2 =
-\frac{1}{2\tilde{\lambda}^2_N}~ e^{m^2/M^2}\left \{\frac{8}{9}
{}~M^2~\frac{\alpha_s}{\pi}~ \int _{0}^{W^2}~ sds~e^{-s/M^2}~(W^2
 + sln\frac{W^2}{s})\right.$$
\be
\left.-\frac{1}{9}~bM^4 E_1 \left (\frac{W^2}{M^2}\right ) + \frac{32}{27}~
\frac{\alpha_s}{\pi}~M^2 a^2 ln \frac{W^2}{M^2} + \frac{8}{9} \pi^2~\Pi M^2 -
\frac{2}{3} m^2_0 a^2 \right \}
\ee
$$\langle \langle V^{NS} \rangle \rangle + A^{NS}_V~M^2 =
-\frac{1}{2\tilde{\lambda}^2_N}~ e^{m^2/M^2}\left \{-\frac{52}{135}
{}~M^2~\frac{\alpha_s}{\pi}~ \int _{0}^{W^2}~ sds~e^{-s/M^2}~(W^2
 + sln\frac{W^2}{s})\right.$$
\be
\left.-\frac{1}{9}~bM^4 E_1 \left (\frac{W^2}{M^2}\right ) - \frac{80}{27}~
\frac{\alpha_s}{\pi}~M^2 a^2 \left (ln \frac{W^2}{M^2}+ 0.9\right )+
\frac{8}{27} \pi^2~R M^2 - \frac{4}{9} m^2_0 a^2 \right \}
\ee
in the
notation(57),(58),(73). Here $\Pi$ and $R$ are unduced by the external field
v.e.v.'s which were estimated by BBK as $\Pi \approx 3.10^{-3}~ GeV^6, R
\sim 1.10^{-3}~ GeV^6$. Using these sum rules, one gets
%90
\be \langle
\langle U^{NS} \rangle \rangle \approx 0.14~ GeV^2
\ee
%91
\be
\langle
\langle V^{NS} \rangle \rangle \approx 0.254~ GeV^2
\ee
The substitution of
(90), (91) into (80) gives the value of twist 4 correction to the Bjorken
sum rule presented above in (10). The examination of the sum rules (88),
(89) shows, however, that the final result (90) comes almost entirely from
the contribution of the last term in OPE in (88) -- the operator of
dimension 8. Recently, A.Oganesian $^{[59]}$ had calculated the contribution
of factorizable v.e.v's of the operators of dimension 10,~ $< 0 \mid
G^2_{\mu \nu} \mid 0 > < \bar{q} q \mid 0 >^2,\\
g^2 < 0 \mid \bar{q}~
\sigma_{\mu \nu}~ G^n_{\mu \nu}~ \lambda^n q \mid 0 >^2$ to the sum rule
(88) and had found that by absolute value it is equal to (90) but has
opposite sign. Of course, we cannot believe in this statement either: it
means only that the results of the calculations are unstable and the value
(10) characteizes the answer by the order of magnitude only.

\vspace{1cm}
\section{Gerasimov, Drell-Hearn (GDH) Sum Rules. The Estimate of Higher
Twist Corrections to $\Gamma_{p,n}$ Using Interpolation between
$\Gamma_{p,n}$ and  GDH Sum Rules.}

\vspace{5mm}

The real photon-nucleon forward scattering amplitude with nucleon spin flip
is expressed through one structure function. In lab.system we can write
%92
\be
e^{(2)}_i(T_{ik})_{spin flip} e^{(1)}_k = i~ \frac{\nu}{m^2}~
\epsilon_{ikl}~ e^{(2)}_i~ e^{(1)}_k~ s_l S_1 (\nu, 0)
\ee
where $e^{(1)}$
and $e^{(2)}$ are polarizations of initial and final photons. At high
energies, $\nu \rightarrow \infty$ according to Regge theory the behaviour
of $S_1(\nu, 0)$ is determined by the exchange of the $a_1$-Regge pole
$^{[17]}$:
%93
\be
S_1(\nu, 0)_{\nu \rightarrow \infty} \sim
\nu^{\alpha_{a_1}(0) - 1}
\ee
Since $\alpha_{a_1}(0) \approx -0.3-0.0$ the
unsubtracted dispersion relation can be written for $S_1(\nu, 0)$

%\end{document}
%94
\be
S_1(\nu) = 4 \int \limits _{0} ^{\infty}~\nu' d\nu'~
\frac{G_1(\nu', 0)}{\nu^{'~2} - \nu^2}
\ee
where $G_1(\nu, Q^2)$ is the
spin-dependent structure function.  Consider the limit $\nu \rightarrow 0$
in (94).  According to the F.Low theorem the terms proportional to $\nu^0$
and $\nu^1$ in the expansion in powers of $\nu$ of the photon-nucleon
scattering amplitude at small $\nu$ are expressed via static characteristics
of nucleon, its charge and anomalous magnetic moment

The calculation gives
%95
\be
S_1(\nu)_{\nu \rightarrow 0} = - \kappa^2,
\ee
where $\kappa$ is the nucleon anomalous magnetic moment: $\kappa_p = 1.79,
\kappa_n = -1.91$.

{}From (94), (95)  the GDH sum rule follows:
%96
\be
\int \limits _{0} ^{\infty} \frac {d \nu}{\nu}~ G_1(\nu, 0) = - \frac{1}{4}
\kappa^2
\ee
Till now no direct check of GDH was done. Only indirect check of (96)was
performed, where in the l.h.s of (96) the parameters of resonances, obtained
from the $\pi N$  scattering phase analysis, where substituted. In this way
with resonances up to 1.8 GeV it was obtained $^{[15, 60]}$  (cf.also
$^{[61]}$)
%97
\be
\begin{array}{lll} &
{}~~~~\mbox{l.h.s of (96)} & ~~~~~~~\mbox{r.h.s. of (96)}\\ \mbox{proton}&
{}~~~~-1.03 & ~~~~~~~ -0.803 \\ \mbox{neutron} & ~~~~-0.83 & ~~~~~~~-0.913\\
\end{array}
\ee
%\end{document}
The l.h.s.  and the r.h.s. of (96) are not in a good agreement -- a
nonresonant contribution is needed. The direct check of the GDH sum rule
would be very desirable!

An important remark: the forward spin dependent photon--nucleon scattering
amplitude has no nucleon pole. This means that there is no nucleon
contribution in the l.h.s. of GDH sum rule -- all contributions come from
excited states: the GDH sum rule is very nontrivial.

In order to connect the GDH sum rule with $\Gamma_{p,n}(Q^2)$  consider the
integrals $^{[17]}$
%98
\be
I_{p,n}(Q^2) = \int^{\infty}_{Q^2/2}~\frac{d\nu}{\nu}G_{1,p,n}(\nu,Q^2)
\ee
It is easy to see that at large $Q^2$
%99
\be
I_{p,n}(Q^2) = \frac{2m^2}{Q^2}~\Gamma_{p,n}(Q^2)
\ee

%\newpage
and at $Q^2=0$ (98) reduces to the GDH sum rule. The $Q^2$ dependence of
$I_{p,n}(Q^2$ is plotted in Fig.11.

\vspace {7cm}
\footnotesize
\begin{center}
Fig.11. ~ The connection o the GDH sum rules with the sum rules at high

$Q^2$ - qualitative $Q^2$ dependence of $I_{p,n}$ and $I_p - I_n$.
\end{center}

\normalsize
\vspace{5mm}

The case of $I_p$ is especially interesting:
$I_p$ is negative at $Q^2=0$ and positive at large $Q^2$,
what indicates to large nonperturbative corrections. In $^{[14]}$ the VDM
based interpolation model was suggested, describing $I_{p,n}(Q^2)$ in the
whole domain of $Q^2$.  The model was improved in $^{[15]}$, where
the contributions of baryonic resonances up to $W=1.8$GeV, taken from
experiment, where accounted. The model has no free parameters, besides the
vector meson mass, for which the value $\mu^2_V=0.6 GeV^2$ was chosen. Using
this model it is possible to calculate the higher twist contributions in
(2),(12). The results are presented in Table 1, as the ratio of asymptotic
$\Gamma^{as}$ with power corrections excluded to the experimentally
measurable $\Gamma$ at given $Q^2$ $^{[39]}$.

%\end{document}
%\newpage

\vspace{5mm}
\centerline{\bf Table 1.}

\vspace{3mm}
\centerline{\underline{Higher twist corrections in GDH sum rule + VDM
inspired model.}}

\vspace{3mm}
\begin{center}
\begin{tabular}{|l|c|c|c|c|}\hline
$Q^2(GeV^2)$  & ~~~~2 & ~~3 & 5 ~~& 10 \\ \hline
$\Gamma^{as}_p/\Gamma_p$ & ~~~~1.44 & ~~1.29 & ~~1.18 & ~~1.08 \\
$\Gamma^{as}_n/\Gamma_n$ & ~~~~1.30 & ~~1.20 & ~~1.13 & ~~1.06 \\
$\Gamma^{as}_{p-n}/\Gamma_{p-n}$ &  ~~~~1.45 & ~~1.29 & ~~1.18 &~ 1.08 \\
$\Gamma^{as}_{p+n}/\Gamma_{p+n}$ &  ~~~~1.47 & ~~1.31 & ~~1.19 & ~~1.08 \\
\hline
\end{tabular}
\end{center}
%\newpage
%\end{document}
%\vspace{5mm}
The power corrections, given in Table 1 are essentially larger (except for
the case of neutron), than the values (10),(86) found in $^{[13]}$.
It must be mentioned that the accuracy of the model in the domain
of intermediate $Q^2$, where it is exploited, is not completely certain.
So, I will consider in what follows the values of the power corrections
presented in Table I as a limiting case of large higher twist corrections.
\newpage
%\vspace{5mm}
\section{Comparison with Experiment.}

When comparing the sum rules (2),(12) with experiment I consider two
limiting variants of perturbative corrections: small with $\Lambda_3=200$MeV
and large with $\Lambda_3=400$MeV. ($\alpha_s$ is computed in 2- loop
approximation, it is assumed that the number of flavours $N_f=3$). For
higher twist correction I also consider two limiting options: small (S),
given by (10),(86) and large (L), determined by the data of Table 1. The
contribution of gluons $\Delta g(Q^2)$ in (12) will be found in the
following way. Let us assume, that at 1 GeV the quark model is valid and
$L_z(1$GeV$^2)=0$  in eq.(18). Taking $\Sigma$=0.3, what is a reasonable
average of the data, we have from (18)
%100
\be
\Delta g (1GeV^2) = 0.35
\ee
The $Q^2$  dependence of $\Delta g$ can be found from the evolution equation
$^{[62]}$
%101
$$ \Delta g(Q^2) = \frac{\alpha_s(\mu^2)}{\alpha_s(Q^2)}\left \{ 1 +
\frac{2N_f}{b\pi}\left [\alpha_s(Q^2) - \alpha_s(\mu^2)\right ] \right \}
\Delta g(\mu^2) + $$
\be
+ \frac{4}{b}\left [ \frac{\alpha_s(\mu^2)}{\alpha_s(Q^2)} - 1 \right ]
\Sigma(\mu^2),
\ee
where $b=11-(2/3)N_f = 9,~ \mu^2 = 1 GeV^2$  and $\Sigma(1 GeV^2) \approx
0.3$. As the calculation shows, the change of scale at which the quark model
is assumed to work (say $0.5 GeV^2$ instead of $1 GeV^2$) or the of use
slightly different $\Sigma(\mu^2)$ in (101)  only weakly influence the
results for $\Sigma$ and $\Delta s$, obtained from experimental data.

I consider the following experimental data (Table 2).

%\newpage
\vspace{5mm}
%\newpage
\centerline{\bf Table 2.}

\vspace{2mm}
\centerline{\underline{The experimental data on $\Gamma_{p,n}$.}}

\vspace{2mm}
\begin{center}
\begin{tabular}{|l|c|l|c|}\hline
$\mbox{Experimental}$  & The target & ~~~~~~~~~~$\Gamma_(p,n)$ &
 \mbox{Mean} $\overline{Q^2} (GeV^2)$ \\
\mbox{group} & & & \\ \hline
EMC [63] & $p$ & $\Gamma_p = 0.126 \pm 0.010 \pm 0.015$ & 10.7 \\
SMC [64] & $p$ & $\Gamma_p = 0.136 \pm 0.011 \pm 0.011$ & 10.5 \\
E143 [65] & $p$ & $\Gamma_p = 0.127 \pm 0.004 \pm 0.010$ & 3 \\
E142 [66] & $He^3$ & $\Gamma_n = -0.022 \pm 0.011$ & 2 \\
SMC [68] & $d$ & $\Gamma_d = 0.034 \pm 0.009 \pm 0.06$ & 10 \\
 & & $\Gamma_p + \Gamma_n = 0.073 \pm 0.022$ & 10 \\
 E143 [67] & $d$ & $\Gamma_d = 0.042 \pm 0.003 \pm 0.004$ & 3\\
 & & $\Gamma_p + \Gamma_n = 0.0908 \pm 0.006 \pm 0.008$& 3 \\
\hline
\end{tabular}
\end{center}

%\newpage
\vspace{5mm}
The last column of Table 2 gives the average $Q^2$ in each experiment.
In
comparison with experiment the perturbative and higher twist corrections, as
well as $\Delta g (Q^2)$ contributions are calculated for these $\overline
{Q^2}$. Experimentally, $Q^2$ are different in different $x$-bins (higher
$Q^2$ at larger $x$). This effect is not accounted in the calculation. The
ratio of $\alpha^3_s$ term to $\alpha^2_s$ term in perturbative corrections
is of order of 1 at $\Lambda_3 = 400 MeV$ and $Q^2 \approx 2-3 GeV^2$ (as
well as $\alpha^4_s/\alpha^3_s$ estimate). For this reason we introduce in
these cases an additional error equal to the $\alpha^3_s$. The values of
$\Sigma$ and $\Delta s$ calculated from comparison of experimental data with
eq.12 are shown in Table 3.(The errors are summed in quadrature).
%\newpage
\newpage
\vspace{5mm}
\centerline{\bf Table 3.}
\vspace{2mm}
\centerline{\underline{Determination of $\Sigma$ and $\Delta s$ from
experimental data}}

\vspace{2mm}
\begin{center}
\begin{tabular}{|c|c|c|c|c|}\hline
Experiment & $\Lambda_3$     & High & $\Sigma$ & $\Delta s$\\
           &~~MeV ~~         & twist &       &            \\ \hline
           & 200            & S    & $0.21 \pm 0.17$ & $-0.13 \pm 0.06$ \\
           \cline{2-4}
 EMC       & 400            & S    & $0.29 \pm 0.17$ & $-0.10 \pm 0.06$ \\
 \cline{2-5}
    p      & 200            & L    &$0.285 \pm 0.17$ & $-0.10 \pm 0.06$\\
 \cline{2-5}
           & 400            & L    & $0.37 \pm 0.17$ & $-0.07 \pm 0.06$ \\
           \hline
           & 200            & S    & $0.30 \pm 0.14$ & $-0.10 \pm 0.05$ \\
           \cline{2-5}
 SMC       & 400            & S    & $0.39 \pm 0.14$ & $-0.07 \pm 0.05$ \\
 \cline{2-5}
  p       & 200            & L    & $0.39 \pm 0.14$ & $-0.07 \pm 0.05$\\
 \cline{2-5}
          & 400            & L    & $0.47 \pm 0.14$ & $-0.04 \pm 0.05$ \\
 \hline
           & 200            & S    & $0.28 \pm 0.10$ & $-0.10 \pm 0.03$  \\
           \cline{2-5}
 E143      & 400            & S    & $0.42 \pm 0.10$ & $-0.06 \pm 0.03$  \\
 \cline{2-5}
    p      & 200            & L    & $0.57 \pm 0.10$ & $-0.006 \pm 0.03$ \\
 \cline{2-5}
           & 400            & L    & $0.71 \pm 0.10$ & $ 0.04 \pm 0.03$ \\
           \hline
 E142      & 200            & S    & $0.60 \pm 0.12$ & $0.003 \pm 0.04$ \\
           \cline{2-5}
  n        & 400            & S    & $0.57 \pm 0.12$ & $-0.005 \pm 0.04$ \\
  \cline{2-5}
 ($^3He$)    & 200            & L    & $0.64 \pm 0.12$ & $0.016 \pm 0.04$ \\
 \cline{2-5}
           & 400            & L    & $0.61 \pm 0.12$ & $0.008 \pm 0.04$ \\
           \hline
           & 200            & S    & $0.27 \pm 0.10$ & $-0.11 \pm 0.03$ \\
           \cline{2-5}
 SMC       & 400            & S    & $0.33 \pm 0.10$ & $-0.09 \pm 0.03$ \\
 \cline{2-5}
 d         & 200            & L    & $0.29 \pm 0.10$ & $-0.10 \pm 0.03$\\
 \cline{2-5}
           & 400            & L    & $0.34 \pm 0.10$ & $-0.08 \pm 0.03$ \\
           \hline
           & 200            & S    & $0.37 \pm 0.06$ & $-0.07 \pm 0.02$ \\
           \cline{2-5}
 E143      & 400            & S    & $0.44 \pm 0.06$ & $-0.05 \pm 0.02$ \\
 \cline{2-5}
     d     & 200            & L    & $0.48 \pm 0.06$ & $-0.04 \pm 0.02$\\
 \cline{2-5}
           & 400            & L    & $0.54 \pm 0.06$ & $-0.015 \pm 0.02$ \\
           \hline
\end{tabular}
\end{center}
\vspace{3mm}
%\newpage
Remark: the contribution to $\Sigma$ of the term proportional to $\Delta g$
is approximately equal to 0.06 in the case of $\Lambda_3 = 200 MeV$ and 0.11
in the case of $\Lambda_3 = 400 MeV$.

%\newpage
If we assume that all the analysed above experiments are correct in the
limits of their quoted errors (or, may be, 1.5 st.deviations), then
requiring  for the results for $\Sigma$ and $\Delta s$ from various
experiments to be consistent, we may reject some theoretical possibilities.
A look at the Table 3 shows that the variant $\Lambda_3 = 400 MeV, L$ (a
contradiction of E143, p  and SMC, d results for $\Sigma$) and, less
certain, the variant $\Lambda_3 = 200 MeV, S$ (a contradiction of E142, n and
SMC, d) may be rejected.

%\newpage

Consider now the Bjorken sum rule. For comparison with theory I choose
combinations of the SMC data - on proton and deuteron, the E143 data - on
proton and deuteron and the E143 data on proton and the E142 on neutron
($^3He$). The results of the comparison of the experimental data with the
theory are given in Table 4.
\newpage

\vspace{5mm}
\centerline{\bf Table 4.}
\vspace{2mm}
\centerline{\underline{Comparison of the experimental data with the Bjorken
sum rule}}

\vspace{2mm}
\begin{center}
\begin{tabular}{|c|c|c|c|c|} \hline
Combination & $(\Gamma_p - \Gamma_n)_{exper.}$ & $\Lambda_3(MeV)$ & High &
$(\Gamma_p - \Gamma_n)_{th}$ \\
of experiments &                             &                   & twist &
\\ \hline
            &                                &    200            & S & 0.193
            \\ \cline{3-5}
  SMC, p    & $0.199 \pm 0.038$              &    400            & S & 0.186
  \\ \cline{3-5}
  SMC, d    &                                &    200            & L & 0.180
  \\ \cline{3-5}
            &                                &    400            & L & 0.173
            \\ \hline
  E143,p          &                          &    200            & S & 0.182
            \\ \cline{3-5}
            & $0.163 \pm 0.010 \pm 0.016$     &    400          & S & 0.168
            \\ \cline{3-5}
  E143,d    &                                &    200            & L & 0.145
            \\ \cline{3-5}

            &                                &    400            & L & 0.134
  \\ \hline
  E143,p &                                   & 200               & S & 0.182
  \\ \cline{3-5}
  E142 n($^3He$) & $0.147 \pm 0.015$          & 400               & S & 0.168
  \\ \cline{3-5}
  recalculated to &                            & 200               & L &
  0.145 \\ \cline{3-5} $\bar{Q}^2 = 3 GeV^2$ &                    & 400
  & L & 0.134 \\ \hline
  \end{tabular}
  \end{center}
  \vspace{3mm}
  From Table 4
we see again some indications for rejection of variants $\Lambda_3 = 400
MeV$, L and, less certain, $\Lambda_3 = 200 MeV$, S.(In the first case there
is a contradiction of the theory with the E143,p and d data, in the second -
with the E143,p , E142, n data).

At existing
experimental accuracy it is impossible to choose from the data the true
values of $\Lambda_3$ and twist-4 correction .
My personal preference is to the variant $\Lambda_3 = 200 MeV$ and to the
value of twist-4 corrections 3 times smaller than given by the GDH sum rule
+ VDM inspired model and, correspondingly, $b_p = 0.04$ in (12), i.e., 2.2
times larger than the BBK result. The argument in the favour of such choice
is that at larger $\Lambda_3$ there will arise many contradictions with the
description of hadronic properties in the framework of the QCD sum rules.
The recent SLAC data $^{[69]}$ on $g_1(x, Q^2)~ Q^2$-dependence
indicate that $\Gamma^{as}_p / \Gamma_p = 1 + c_p/Q^2, c_p = 0.25 \pm 0.15$
what is compatible with the estimate above. In this case all experimental
data except for E142,n, are in a good agreement with one another and the
values of $\Sigma$ and $\Delta s$ averaged over all experiments, except for
E142,n are
%102
\be
\overline{\Sigma} = 0.35 \pm 0.05 ~~~~~~~~~~~ \Delta s =
-0.08 \pm 0.02
\ee
(see also $^{[70]}$ where the values close to (102) were
obtained).

The values of $\Sigma$ and $\Delta s$ obtained from the E142,n experiment at
such a choice of $\Lambda_3$ and twist-4 corrections, are different:
%103
\be
\Sigma = 0.61 \pm 0.12 ~~~~~~~~~~ \Delta s = 0.01 \pm 0.04
\ee
It is impossible to compete (102),(103) by any choice of $\Lambda_3$ and
higher twist correction. Perhaps, this difference is caused by inaccounted
systematic errors in the E142 experiment.

%\newpage

A remarkable feature of the result (102) (as well as of the data in Table 3)
is the large value of $\mid \Delta s \mid$ - the part of the proton spin
projection carried by strange quarks. This value may be compared with the
part of the proton momentum carried by strange quarks
%104
\be
V^s_2 = \int~ dx ~ x \left [ s_+(x) + s_-(x)\right ] = 0.026 \pm
0.006~^{[71]}, ~~~~ 0.040 \pm 0.005~^{[72]}
\ee
\newpage
The much larger value of
$\mid \Delta s \mid$ in comparison with $V^s_2$ contradicts the standard
parametrization
%105
$$s_+(x) + s_-(x) = A~x^{-\alpha}~ (1 -
x)^{\beta}$$
\be
s_+(x) - s_-(x) = B~x^{-\gamma}~(1-x)^{\beta}
\ee
and requirement of
positiveness of $s_+$ and $s_-$, if $\alpha \approx 1$(pomeron intercept) and
$\gamma \leq 0$ ($a_1$ intercept). Large $\mid \Delta s \mid$ and small
$V^s_2$ means that the transitions $\bar{s} s \rightarrow \bar{u} u +
\bar{d} d$ are allowed in the case of the operator $j_{\mu 5}$ and are
suppressed in the case of the quark energy-momentum tensor operator
$\Theta_{\mu \nu}$, corresponding to the matrix element $V_2$. Such
situation can be due to nonperturbative effects and the instanton
mechanism for its explanation was suggested $^{[73]}$. It must be
mentioned that in the suggested by Brodsky $^{[74]}$ more refined
parametrization, which takes into account the fact that at $x
\rightarrow 1~~~ q_-(x)/q_+(x) \sim (1 - x)^2$, the contradiction
weakens. Improvement of experimental accuracy is necessary in order to be
sure that the inequality \\
$\mid \Delta s \mid \gg V^s_2$ indeed takes
place.

\vspace{5mm}
\section{The calculations of the polarized structure functions by QCD
sum rule method}

\vspace{2mm}

I recall the basing points of the calculation of nucleon structure functions
in the QCD sum rule approach (for details see $^{[75]}$. Consider four-point
correlator
%106
$$
T^{\pm}_{\mu \nu} (p, q) = -i~ \int ~ d^4 xd^4 y d^4 z~ e^{iqx}~ e^{ip(y-z)}
\times$$
\be
\times < 0 \mid T~\left \{\eta(y),~ j^{\mp}_{\mu}(x),~j^{\pm}_{\nu}(0),~
\bar{\eta}(z)\right \} \mid 0 >
\ee
where
$$
j^+_{\mu} = \bar{u} \gamma_{\mu} (1 + \gamma_5) d, ~~~ j^-_{\mu} = \bar{d}
\gamma_{\mu} (1 + \gamma_5) u
$$
and $\eta$ is given by (36). (In order to separate $u$- and $d$ quark
distributions it is convenient to consider $W^{\pm}$ proton scattering).
Suppose that $q^2 < 0, p^2 < 0, q^2 = -Q^2,~~ Q^2$ is large
enough, $Q^2 \sim 10~GeV^2, Q^2 >> \mid p^2 \mid$ and retain only the
leading terms in the expansion over $p^2/Q^2$. (This corresponds to account
of only twist 2 contributions). Assume also that $\mid p^2 \mid >>
R^{-2}_c$, where $R_c$ is the confinement radius and perform OPE in $1/p^2$
for the discontinuity of $T_{\mu \nu}$ in the $s$-channel
%107
\be
Im~
T^{\pm}_{\mu \nu} = \frac{1}{2i}~ \left [T^{\pm}_{\mu \nu} (p^2, q^2, s + i
\epsilon) - T^{\pm}_{\mu \nu} (p^2, q^2, s - i \epsilon)\right ]
\ee
At the
first sight it seems, that OPE is not legitimate here because $T_{\mu \nu}$
is the forward scattering amplitude in which large distances in the
$t$-channel are of importance. This, indeed, would be the case if we would
consider the exclusive object $Re~ T_{\mu \nu}$. But for $Im~ T_{\mu \nu}$
which in fact is an inclusive observable, the situation is completely
different. Consider first the bare loop diagram for $Im~ T_{\mu \nu}$ (the
contribution of dimension $0$ operator in OPE) -- Fig.12.
\newpage
\vspace*{3.5cm}
\footnotesize
\begin{center}
Fig.12 The bare loop diagram for $Im~ T_{\mu \nu}$. The crosses mean on
mass shell propagators.
\end{center}
\vspace{5mm}
\normalsize
%\normalsize
The direct calculation of the diagram gives that the virtuality of the
active quark on which the scattering proceeds is equal to $(x = Q^2/2 \nu,~
\nu = pq)$
%108
\be
k^2 = p^2x - \frac {k^2_{\bot}}{1 - x}
\ee
and $k^2_{\bot}$ is of order $k^2_{\bot} \sim \mid p^2 \mid x (1-x)$.
Therefore, $k^2 \sim p^2 x$. (Strictly speaking, this statement refers only
to terms in the amplitude, singular in $p^2$, but just these terms are of
interest for us since the Borel transformation in $p^2$ killing the regular
in $p^2$ terms will be performed later). Since it was assumed that $\mid p^2
\mid >> R^{-2}_c$ we come to a conclusion that at not small $x, \mid k^2
\mid$ and $k^2_{\bot}$ are large and OPE is legitimate.

A general proof of the above mentioned statement follows from the fact that
at large $\mid p^2 \mid$, $\mid q^2 \mid$ the nearest to zero singularity in
$t$ of the function $Im~T(p^2, q^2, s, t)$ is determined by the boundary of
the Mandelstam spectral function which is found to be
%109
\be
t =
4~\frac{p^2 q^2}{s} = - 4~ \frac{x}{1-x}~p^2
\ee
So, at $t = 0$, large $\mid
p^2 \mid$ and not small $x$, we are far from the boundary of the spectral
function in $t$-channel and OPE is valid.

The statement that the method does not work at small $x$ is evident
beforehand: this is the Regge domain, where OPE cannot work.

The estimate of active quark virtualities (108) have much more general
 meaning, beyond QCD sum rule appoach. In lepton-hadron scattering $\mid p^2
 \mid$ is of order of QCD scale, $\mid p^2 \mid \sim 1~ GeV^2$. Then at small
$x~~ \mid k^2 \mid$ are small and we are in the nonperturbative domain of
 QCD. This means that the interaction of quarks with nonperturbative vacuum
  fields, especially gluonic fields, are important. The direct calculation in
 QCD sum rule approach supports this expectation.

 The method in view is also invalid at $x$ close to $1, ~ 1 - x << 1$. This
  is evident, because it is a resonance region. Finally, we restrict
  ourselves to intermediate $Q^2 \sim 5-10~ GeV^2$, since the evolution of
 the structure functions will not be accounted.

 The calculation proceeds in the standard way of the QCD sum rule method. In
  the QCD side of the sum rule $Im~T^{\pm}_{\mu \nu}$ is calculated by OPE
 with the account of v.e.v.of various operators. For the nonpolarized case
 the gluonic condensate and the term $\sim \alpha_s < 0 \mid \bar{q}q\mid 0
 >^2$ were accounted $^{[75]}$. For polarized structure
functions $g_1$ and $g_2$, besides the bare loop, only the
contribution  of the term $\sim \alpha_s < 0 \mid \bar{q}q\mid 0
 >^2$ was calculated $^{[76]}$, since it is expected that at
intermediate $x$, where the results are correct, the
contibution of gluonic condensate is small. The v.e.v.$< 0 \mid \bar{q}q\mid
 0 >^2$ do not contribute to the sum rule, because it is concentrated at $x =
1$ (proportional to $\delta(1-x)$ in twist 2 terms) -- outside of the
applicability domain of the method. (Only chirality conserving structure were
considered in $^{[75,76]}$). The hadronic side of the sum rule is represented
schematically by Fig.13.
\newpage
\vspace*{3cm}
\footnotesize
\begin{center}

Fig.13. The representation of hadronic part of the sum for determination of
the nucleon structure functions:
a) the double nucleon pole term which gives the desired structure function;
b) the background term, when in the course of the deep inelastic scattering
$N$ undergoes transition to excited state $N^*$;
c) continuum.
\end{center}
\vspace{5mm}
\normalsize
The Borel transformation is applied to both sides of the sum rule. The
continuum contribution is suppressed by the Borel transformation, it is
approximated by the bare loop (Fig.12) and transferred to the QCD side. The
background term -- Fig.13b -- in its dependence on the Borel parameter $M^2$
differs from the nucleon term by an additional factor $M^2$.This
circumstance permits one to kill this term in the same way, as it was killed
the constant $A$ in (56).

In the case of polarized structure functions $g_1$ and $g_2$ it was
shown $^{[76]}$ that for the bare quark loop the Bjorken and
Burkhardt-Cottingham $^{[77]}$ sum rules are fulilled. It was found that for
 the function $g_1(x)$ the results are reliable in a rather narrow domain of
 intermediate $x$:  $0.5 \leq x \leq 0.7$.In this domain the contribution of
 $u$-quarks $g^u_1$ is much larger than $d$-quarks, $g^u_1 \gg g^d_1$.
 Therefore, $g_1 \approx (4/9)~g^u_1$.  $g^u_1$ was calculated and for the
 mean value of $g_1$ in this interval it was obtained (at $Q^2 \sim 5-10
 GeV^2$):
 %110
 \be
 \bar{g}_1 (0.5 < x < 0.7) = 0.05 \pm 50\% \
\ee
The large
 uncertainty in (110) results from the large contribution of nonleading term
in OPE and from large background at the phenomenological side of the QCD sum
rule. The E143 proton $^{[65]}$ and deuteron $^{[68]}$ data (the latter
under assumption that $g^n_1$ is small in this interval of $x$) give roughly
the same values:
%111
\be
\bar{g}_1~ (0.5 \leq x \leq 0.7) = 0.08 \pm 0.02
\ee
This value is in a good agreement with the SMC result $^{[64]}$
%112
\be
\bar{g}_1 (0.4 < x < 0.7) = 0.08 \pm 0.02 \pm 0.01
\ee
and
with
theoretical expectation (110).

For  the case of the structure function $g_2$ only $g^u_2$ can be calculated
at $0.5 < x < 0.8$, the calculation of $g^d_2$ fails because of large
contribution of nonleading terms in OPE. If we assume that like in the case
of $g_1,~\mid g^d_2 \mid \ll \mid g^u_2 \mid$, then $^{[76]}$
%113
\be
g_2 (0.5 < x < 0.8) = -0.05 \pm 50\%
\ee
The E143 data $^{[78]}$ in this interval of $x$ are:
%114
\be
g_2(0.5 < x < 0.8) = -0.037 \pm 0.020 \pm 0.003
\ee
in a good agreement with (113)

The serious disadvantages of the QCD sum rules calculations of the structure
functions are:

1) large contribution of nonleading terms in OPE;

2) large background terms from inelastic transitions $N \rightarrow N^*$
(Fig.13b) at the physical side of the sum rules. In order to kill these
terms we are forced to differentiate the sum rules over the Borel parameter
$M^2$. This operation increases the role of nonleading terms in OPE and
continuum and deteriorates the accuracy of the sum rule. A possible way to
overcome this drawback of the method is to start from nonforward scattering
amplitudes $q_1 \not= q_2, p^2_1 \not= p^2_2$ and to use the double Borel
transformation in $p^2_1$ and $p^2_2$. The calculations in such approach will
be much more complicated, but, may be the game is worth of candles.

\section{Calculation of Chirality Violating Structure Function $h_1(x)$ by
QCD Sum Rules}
\vspace{3mm}
%%%%%%%%%%%%%%%%%%%%%%%%%%%%%%%%%%%%%%%%%%%%%%%%%%%%%%%%%%%%%%%%%%%%%%%%%%

As is well known, all structure functions of the twist two -
$F_1(x), ~F_2(x), ~g_1(x)$, which are measured in the deep-inelastic
lepton-nucleon scattering, conserve
chirality. Ralston and Soper $^{79}$ first
demonstrated that besides these structure functions, there exists the
twist-two chirality violating nucleon structure function $h_1(x)$.
This structure function does not manifest itself in the deep inelastic
lepton-hadron scattering, but can be measured in the Drell-Yan process with
both beam and target transversally polarized. The reason of this circumstance
is the following. The cross section of the deep inelastic electron(muon)-hadron
scattering is proportional to the imaginary part of the forward virtual
photon-hadron scattering amplitude.\\
\vspace{5mm}

\footnotesize
\noindent
\begin{tabular}{p{75mm}p{75mm}}
 & \underline{\bf Fig.14}\\
 & a) Deep inelastic lepton-hadron scattering, the quark chiralities are
conserved. Solid lines are quarks, wavy lines are virtual photons, $R(L)$
denote right (left) chirality of quarks; b) Drell-Yan process with
chirality of quarks flipped.
\end{tabular}

\vspace{5mm}
%\noindent
\normalsize
At high photon virtuality the quark
Compton amplitude dominates, where the photon is absorbed and emitted by the
same quark (Fig.14a) and the conservation of chirality is evident.
The  cross section of the Drell-Yan process can be represented as an
imaginary part of the diagram, Fig.14b. Here virtual photons interact with
different quarks and it is possible, as  is  shown  in
Fig.14b, that chirality violating amplitude in Drell-Yan processes is not
suppressed at high $Q^2$ in comparison with chirality conserving ones,
and consequently, corresponds to twist two. This amplitude,
corresponding to target spin flip, has no parton interpretation in terms
of quark distritutions in the helicity basis and, as was shown by Jaffe
and Ji $^{[80]}$ can be only represented as an element of the quark-quark
density matrix in this basis.

However, $h_1(x)$ can be interpeted $^{80}$ as a difference of quark
densities with the eigenvalues $+1/2$ and $-1/2$ of the transverse
Paili-Lubanski spin operator $\hat s_{\perp} \gamma_5$ in the transversely
polarized proton. It this basis $h_1(x)$ can be described in terms of
standard parton language.

The proton structure function $h_1(x)$ can be defined in the light cone
formalism as follows $^{80}$
%  115
\be
   \begin{array}{cc}
i \int \frac{d \lambda}{2 \pi} e^{i \lambda x} < p, s \mid \bar{\psi}
(0) \sigma_{\mu \nu} \gamma_5 \psi (\lambda n) \mid p, s > =
2 [ h_1 (x, Q^2)(s_{\perp \mu} p_{\nu} - s_{\perp \nu} p_{\mu})
+ \\
   + h_L (x, Q^2) m^2 (p_{\mu} n_{\nu} - p_{\nu} n_{\mu}) (s n) +
   h_3 (x, Q^2) m^2 (s_{\perp \mu} n_{\nu} - s_{\perp \nu} n_{\mu}) ] ~.
   \end{array}
\ee
Here $n$ is a light cone vector of dimension (mass)$^{-1}, n^2 = 0,
n^+ = 0, pn = 1, p$ and $s$ are the proton momentum and spin vectors,
$p^2 = m^2, s^2 = -1, ps = 0$ and $s = (sn)p + (sp)n + s_{\perp},
h_{L}(x, Q^2)$ and $h_3 (x, Q^2)$ are twist-3 and 4 structure functions.
For comparison in the same light cone notation the standard structure
function $F_1(x, Q^2)$ is given by
%  116
\be
   \int \frac{d \lambda}{2 \pi} e^{i \lambda x} < p, s \mid \bar{\psi}
   (0) \gamma_{\mu }  \psi (\lambda n) \mid p, s > =
   4 [F_1 (x, Q^2) p_{\mu} + M^2 f_4 (x, Q^2) n_{\mu} ]
\ee
(Eqs. (115), (116) are written for one flavour).

Basing on the definitions (115),(116) an inequality was proved in $^{80}$
%  117
\be
   q (x) ~ \ge ~ h_1 ^q (x) ~ ,
\ee
which holds for each flavour $q = u, d, s$. (Here $h_1 ^q (x)$ is the
flavour $q$ contribution to $h_1 (x)$.)  Recently, Soffer $^{[81]}$ (see
also $^{[82]}$ derived an inequality
% 118
\be
\mid h^u_1 \mid < \left [u(x) + g^u_1(x)\right ] /2
\ee

$h_1 (x)$ can be also represented through a $T$-product of currents $^{[83]}$
% 119
\be
   T_{\mu}(p, q, s) = i \int d^4 x e^{iqx} < p, s \mid (1/2) T
   \{ j_{\mu 5} (x), j (0) + j (x), j_{\mu 5} (0) \} \mid p, s > ~ ,
\ee
where $j_{\mu 5} (x)$ and $j(x)$ are axial and scalar currents.

The general form of $T_{\mu}(p, q, s)$ is
% 120
$$  T_{\mu}(p, q, s) =  \left( s_{\mu} - \frac{q s}{q^2} q_{\mu} \right)
    \tilde h_1 (x, Q^2) + \left( p_{\mu} - \frac{\nu q_{\mu}}{q^2} \right)
    (q s ) l_1 (x, Q^2) ~ +
$$
\be
   + ~ \varepsilon_{\mu \nu \lambda \sigma} p_{\nu} q_{\lambda}s_{\sigma}
   (q s) l_2 (x,q^2)
\ee
(only spin--dependent terms are retained). It can be proved $^{83}$, that
% 121
\be
   h_1 (x, Q^2) ~ = ~ -\frac{1}{\pi} Im \tilde h_1 (x, Q^2) ~ .
\ee
As is clear from (115) or (119) $h_1(x)$ indeed violates chirality.

Since $h_1(x)$ violates chirality one may expect that in QCD it can be
expressed in terms of chirality violating fundamental parameters of the
theory, the simplest of which (of the lowest dimension) is the quark
condensate. Basing on this idea $h_1(x)$ calculation was performed
$^{[83]}$.  The main difference in comparison with the structure functions
considered in Sec.6 is that in the case of $h_1(x)$ determination the
chirality violating structure is studied. As a result, $h_1(x)$ was found to
be proportional to the quark condensate $< 0 \mid \bar{u} u \mid 0
>$ with the correction term in OPE proportional to the mixed
quark-gluon condensate $g < 0 \mid \bar{u} \sigma_{\mu
\nu}~G^n_{\mu \nu}~\lambda^n u \mid 0 >$. It was obtained that
for proton $h^u_1(x) \gg h^d_1(x)$ and as a consequence $h_1(x) \approx
(4/9) h^u_1(x)$. The calculations in $^{[83]}$ are valid at $0.3 < x <
0.6$.The extrapolation in the region of small $x$ can be performed using
the Regge behaviour $h_1(x) \sim x^{-\alpha_{a_1}}$, the extrapolation in
the region o large $x$, using the inequalities (117),(118). The final result
of the calculation of $h^u_1(x)$ with the extrapolation is shown in Fig.15.
\newpage
\vspace*{7cm}
%\footnotesize
\begin{center}
Fig.15
\footnotesize
The $u$-quark contribution to the proton structure

function $h_1(x)$ based on the QCD sum rule calculation

$^{[83]}$ at intermediate $x$,~ $0.3 < x < 0.6$. At $x < 0.3$

an extrapolation according to Regge behaviour was performed.

The dashed line represents the Soffer inequality (118).
\end{center}
\vspace{5mm}
\normalsize
It is expected that the accuracy  of $h^u_1(x)$ determination is
about 30\% at $x = 0.4$ and about 50\% at $x = 0.6$. The inequality
$h^u_1(x) > g^u_1(x)$ suggested in $^{[80]}$ was confirmed.  Numerically,
$h_1(x)$ is rather large, that gives a good chance for its experimental
study.

\vspace{5mm}

\section{Conclusions}

\vspace{3mm}
%\normalsize
The experimental study of the nucleon spin structure, where very impressive
results were obtained, triggered many theoretical investigations. As
result, we understand now much more about the internal content of nucleon
and, even generally, about the structure of QCD. We know, that Ellis-Jaffe
sum rule is not the last word in the problem of the nucleon spin content:
gluons and strange quarks are of importance in this problem. The connection
of gluon contribution to the nucleon spin to the anomaly is clarified. More
clear becomes the role of nonperturbative phenomena in QCD.  In this aspect
there is a very important indication that the part of the proton spin
projection carried by strange quarks $\mid \Delta s \mid$ is much larger
than the part of the proton momentum $V^s_2$ carried by strange quarks,
$\mid \Delta s \mid \gg V^s_2$. (A confirmation of this with a better
accuracy is necessary!). If confirmed, this statement indicates a nontrivial
dynamics of QCD vacuum and its explanation is a challenge for a theory. It
would be very desirable if experimentally it would be possible in the near
future:\\

1. To increase the accuracy by 2-3 times.

2. To study the $Q^2$-dependence of $g_1(x, Q^2)$(in separate bins in $x$).

3. To go to higher $Q^2$: probably this can be done at HERA. (The experiment
at higher $Q^2$ will be informative if only its accuracy will not be worse
than the existing ones).

4. To have better data in the domain of small $x$ -- probably, this also can
be done at HERA.

5. To perform measurements of two-jets events in polarized deep-inelastic
scattering, which will give information about $\Delta g$.

6. To have more data on $s$-quark distribution in nonpolarized nucleon,
expecially at $x < 0.1$.

7. To perform new experiments on elastic $\nu p$-scattering from which one
can find the combination $\Delta u - \Delta d - \Delta s$, as well as
elastic $ep$ scattering with separating the $Z$-exchange term.

8. To have better data on $g_2$.

9. To measure $h_1(x)$.

10. To perform a direct check of the GDH sum rule.

\vspace{2mm}
The problem confronting the theory, which, I believe, could be solved in the
near future, are:

1. The lattice calculation determination of the induced by external field
$v.e.v's$, especially those which cannot be calculated by the QCD sum rule
approach.

2. The study of $\bar{s}s$ and $\bar{u} u + \bar{d} d$ mixing in the nucleon
(on lattice and in the instanton liquid model), especially for the case of
the axial and energy-momentum tensor operators.

3. Improvement of the calculation of the structure functions by the QCD sum
rule method.

4. To achieve a better understanding of nonperturbative effects in the
structure functions at small (but not very small) $x \sim 10^{-2}$.

\vspace{5mm}
\begin{center}
\underline{Acknowledgement}
\end{center}

\vspace{3mm}

I am thankful to B.Frois and V.Hughes or their invitation to participate at
Erice School, for useful discussion and hospitality at Erice. This
investigation was supported in part by the International Science Foundation
Grant M9H300 and by International Association for the Promotion of
Cooperation with scientists from Independent States o the Former Soviet
Union Grant INTAS-93-0283.

%\end{document}


\newpage
\begin{thebibliography}{999}
\bibitem{1} J.D.Bjorken, Phys.Rev. {\bf 148}, (1966) 1467.
\bibitem{2} J.Kadaira et al., Phys.Rev. {\bf D20} (1979) 627, Nucl.Phys.
{\bf B159} (1979) 99, {\bf B165} (1980) 129.
\bibitem{3} S.A.Larin and J.A.M.Vermaseren, Phys.Lett.{\bf B259}
(1991) 345.
\bibitem{4} A.L.Kataev, Phys.Lett.{\bf D50} (1994) 5469.
\bibitem{5} M.Aguilar--Benitez et al.,Particle Data Group,Phys.Rev.
{\bf D50} (1994) 1173.
\bibitem{6} P.B.Mackenzie and C.P.Lepage, Phys.Rev.Lett.{\bf 47}
(1981) 1244.
\bibitem{7} M.Gl\"uck, E.Reya and A.Vogt, Zs.Phys.{\bf C53} (1992)
127.
\bibitem{8} K.Abe et al., SLD Collab., preprint SLAC-PUB-6739 (1995).
\bibitem{9} R.Akers et al., OPAL Collab., preprint CERN-PPE/95-069 (1995).
\bibitem{10} Y.K.Li et al., AMY Collab., Phys.Lett.{\bf B355} (1995)
394.
\bibitem{11} A.Pich, Invited talk at QCD 94 Workshop, Monpellier, July
(1994), FTUV/94-71 (1994).
\bibitem{12} A.Shifman, preprint TPI-MINN-94/42-T (1994)

S.Neubert, preprint CERN-TH 7524/94 (1995).
\bibitem{13} I.I.Balitsky, V.M.Braun and A.V.Kolesnichenko, Phys.Lett.
{\bf B242} (1990) 245, Errata {\bf B318} (1993) 648.
\bibitem{14} M.Anselmino, B.L.Ioffe and E.Leader, Sov.J.Nucl.Phys.
{\bf 49} (1989) 136.

\bibitem{15} V.D.Burkert and B.L.Ioffe, ZheTF {\bf 105} (1994) 1153.
\bibitem{16} I.G.Aznauryan, Phys.Atom.Nucl. {\bf 58} (1995) 1014.
\bibitem{17} B.L.Ioffe, V.A.Khoze and L.N.Lipatov, Hard Processes, v.I North
Holland, Amsterdam, 1984.
\bibitem{18} S.Gerasimov, Yad.Fiz. {\bf 2} (1966) 930.

S.D.Drell and A.C.Hearn, Phys.Rev.Lett.{\bf 16} (1966) 908.
\bibitem{19} S.A.Larin, Phys.Lett.{\bf B334} (1994) 192.
\bibitem{20} M.A.Ahmed, G.G.Ross,, Nucl.Phys.{\bf B111} (1976) 441.
\bibitem{21} C.S.Lam, Bing-An Li, Phys.Rev. {\bf D25} (1982) 683.
\bibitem{22} A.V.Efremov, O.V.Teryaev, Dubna Preprint JINR-E2-88-287 (1988).
\bibitem{23} G.Altarelli, G.G.Ross, Phys.Lett. {\bf B212} (1988) 391.
\bibitem{24} R.D.Carlitz, J.C.Collins, A.H.Mueller, Phys.Lett. {\bf B214}
(1988) 229.
\bibitem{25} R.L.Jaffe, A.Manohar, Nucl.Phys. {\bf B337} (1990) 509.
\bibitem{26} S.Forte, Phys.Lett. {\bf B224} (1989) 189; Nucl.Phys. {\bf
B331} (1990) 1.
\bibitem {27} G.Altarelli, W.J.Stirling, Particle World {\bf 1} (1989) 40.
\bibitem{28} G.Veneziano, Mod.Phys.Lett. {\bf A4} (1989) 1605.

G.M.Shore, G.Veneziano, Phys.Lett. {\bf B244} (1990) 75; Nucl.Phys.{\bf
B381} (1992) 23.  \bibitem{29} G.Altarelli, B.Lampe, Z.Phys. {\bf C47}
(1990) 315.  \bibitem{30} G.T.Bodwin, J.Qiu, Phys.Rev. {\bf D41} (1990)
2755.  \bibitem{31} L.Mankiewicz, A.Sch\"afer, Phys.Lett. {\bf B242} (1990)
455;

L.Mankiewicz, Phys.Rev. {\bf D43} (1991) 64.
\bibitem{32} T.P.Cheng, L.F.Li, Phys.Rev.Lett. {\bf ~ 62} (1989) 1441;

Carnegie-Mellon Univ.Preprint CMU-HEP-90-2 (1990).
\bibitem{33} A.V.Manohar, Phys.Rev.Lett. {\bf~ 66} (1991) 289.
\bibitem{34} W.Vogelsang, Z.Phys. {\bf C50} (1991) 275.
\bibitem{35} S.D.Bass, B.L.Ioffe, N.N.Nikolaev, A.W.Thomas, J.Moscow Phys.Soc.
{\bf 1} (1991) 317.
    \bibitem{36} E.Reya, in: Proc. of Intern.Workshop "QCD-20 Years Later",

Aachen, 1992 (World Scientific, 1993).
\bibitem{37} S.L.Adler, Phys.Rev. {\bf 177} (1969) 2426.

J.Bell and R.Jackiw Nuovo Cim. {\bf A51} (1969) 47.
\bibitem{38} S.Y.Hsueh et al., Phys.Rev. {\bf D38} (1988) 2056.
\bibitem{39} B.L.Ioffe, Surveys in High Energy Physics, {\bf 8} (1995) 107.
\bibitem{40} J.Lichtenstadt and H.Lipkin, preprint TAUP-2244-95 (1995).
\bibitem{41} V.M.Belyaev, B.L.Ioffe and Ya.I.Kogan, Phys.Lett.
{\bf 151B} (1985) 290.
\bibitem{42} J.Ellis and R.L.Jaffe, Phys.Rev.{\bf D9}(1974),
{\bf D10} (1974) 1669.
\bibitem{43} S.J.Brodsky, J.Ellis and M.Karliner, Phys.Lett.{\bf B20}
(1988) 309.

J.Ellis and M.Karliner, Phys.Lett.{\bf B21} (1988) 73.
\bibitem{44} Z.Ryzak, Phys.Lett.{\bf B217} (1989) 325.

V.Bernard, U.G.Meissner Phys.Lett. {\bf B223} (1989) 439.
\bibitem{45} V.M.Belyaev and Ya.I.Kogan, JETP Lett. {\bf 37} (1983) 730.
\bibitem{46} B.L.Ioffe and A.V.Smilga, Pisma v ZhETF {\bf 37} (1983)
250,

Nucl.Phys.{\bf B232} (1984) 109.
\bibitem{47} I.I.Balitsky and A.V.Yung, Phys.Lett.{\bf B129} (1983)
328.
\bibitem{48} V.A.Novikov et al., Nucl.Phys. {\bf B237} (1984) 525.
\bibitem{49} B.L.Ioffe, Phys.Atom.Nucl. {\bf 58} (1995) 1408.
\bibitem{50} B.L.Ioffe, Nucl.Phys. {\bf B188} (1981) 317,
{\bf B191} (1981) 591 E.

V.M.Belyaev and B.L.Ioffe, Sov.Phys.JETP {\bf 56} (1982) 493.
\bibitem{51} B.L.Ioffe and A.Yu.Khodjamirian, Yad.Fiz.{\bf 55} (1992) 3045

\bibitem{52} V.A.Novikov et al. Phys.Lett. {\bf 86B} (1979) 347.
\bibitem{53} B.V.Geshkenbein and B.L.Ioffe, Nucl.Phys. {\bf B166}
(1980) 340.
\bibitem{54} V.A.Novikov, M.A.Shifman, A.I.Vainstein and V.I.Zakharov,
Nucl.Phys. {\bf B191} (1981) 301.
\bibitem{55} E.V.Shuryak, Rev.Mod.Phys. {\bf 65} (1993) 1.
\bibitem{56} M.Fukigita et al., Phys.Rev. {\bf D51} (1995) 3952.
\bibitem{57} M.Anselmino, A.Efremov and E.Leader, preprint CERN-TH~7216/94.
\bibitem{58} E.V.Shuryak and A.I.Vainstein, Nucl.Phys.
{\bf B201} (1982) 144.
\bibitem{59} A.Oganesian, ITEP preprint, in preparation.
\bibitem{60} V.D.Burkert and Zhujun Li, Phys.Rev.{\bf D47} (1993) 46.
\bibitem{61} R.L.Workman and R.A.Arndt, Phys.Rev.{\bf D45} (1992)
1789.
\bibitem{62} M.Gl\"uck and E.Reya, Phys.Lett. {\bf B270} (1991) 65.
\bibitem{63} J.Ashman et al., Nucl.Phys.{\bf B238} (1989) 1.
\bibitem{64} D.Adams et al., Phys.Lett. {\bf B329} (1994) 399.
\bibitem{65} K.Abe et al., Phys.Rev.Lett.{\bf 74} (1995) 346.
\bibitem{66} P.L.Anthony et al., Phys.Rev.Lett. {\bf 71} (1993) 959.
\bibitem{67} D.Adams et al., preprint CERN-PPE/95-97 (1995).
\bibitem{68} K.Abe et al., Phys.Rev.Lett {\bf 75} (1995) 25.
\bibitem{69} K.Abe et al., E143 Collab., preprint SLAC-PUB-95-6997 (1995).
\bibitem{70} J.Ellis and M.Karliner, preprint CERN-TH-7324/94
\bibitem{71} H.Abramowicz et al., Zs.Phys.{\bf C15} (1982) 19.
\bibitem{72} A.O.Bazarko et al., CCFR Collab., Zs.Phys. {\bf C65} (1995)
189.
\bibitem{73} B.L.Ioffe and M.Karliner, Phys.Lett.{\bf 247B} (1990)
387.
\bibitem{74} S.J.Brodsky, in: Lecture at SLAC Summer Institute on Particle
Physics, 26 July - 6 August, 1993, preprint SLAC-PUB-6450, 1994.
\bibitem{75} V.M.Belyaev  and B.L.Ioffe, Nucl.Phys. {\bf B310} (1988) 548.
\bibitem{76} V.M.Belyaev and B.L.Ioffe, Intern.Journ.of Mod.Phys.
{\bf A6} (1991) 1533.
\bibitem{77} H.Burkhardt and W.H.Cottigham, Ann. of Phys. {\bf 56} (1970)
453.
\bibitem{78} K.Abe et al., preprint SLAC-PUB-6982 (1995).
\bibitem{79} J.Ralston and D.E.Soper, Nucl.Phys. {\bf B152} (1979) 109.
\bibitem{80} R.L.Jaffe and X.Ji, Phys.Rev.Lett. {\bf 67} (1991) 552,
 Nucl.Phys. {\bf B375} (1992) 527.
\bibitem{81} J.Soffer, Marseille  prepint  CPT-94/p.3059  (1994).
\bibitem{82} G.R.Goldstein, R.L.Jaffe and X.Ji, preprint MIT-CTP-2402
(1995).
\bibitem{83} B.L.Ioffe and A.Yu.Khodjiamirian, Phys.Rev. {\bf D51} (1995)
3373.

\end{thebibliography}
\end{document}